\documentclass[trackchanges,twocolumn]{aastex62}
\pdfoutput=1
\usepackage{natbib}
\usepackage{amssymb,amsmath}
\usepackage{graphicx}
\usepackage{color}

\newcommand{\uofa}{\affiliation{ Lunar and Planetary Laboratory, University of Arizona, Tucson, AZ 85721, USA}}
\newcommand{\uofap}{\affiliation{Department of Physics, University of Arizona, Tucson, AZ 85721, USA}}

\usepackage{fancyhdr}

\usepackage[T1]{fontenc}

\usepackage{float}

\UseRawInputEncoding

\newcommand{\revise}[1]{{#1}}

\begin{document}

\title{\revise{Estimation of turbulent proton and electron heating rates via Landau damping constrained by Parker Solar Probe observations}}


\author[0000-0002-8941-3463]{Niranjana Shankarappa}\uofap
\author[0000-0001-6038-1923]{Kristopher G. Klein}\uofa
\author[0000-0002-7365-0472]{Mihailo M. Martinovi\'c}\uofa


\begin{abstract}
The heating of ions and electrons due to turbulent dissipation plays a crucial role in the thermodynamics of the solar wind and other plasma environments. Using magnetic field and thermal plasma observations from the first two perihelia of the Parker Solar Probe (PSP), we model the relative heating rates as a function of radial distance, \revise{magnetic spectra, and plasma conditions,} enabling us to better characterize the thermodynamics of the inner heliosphere. \revise{We employ the Howes et al. 2008 steady-state cascade model, which considers the behavior of turbulent, low-frequency, wavevector-anisotropic, critically balanced Alfv\'enic fluctuations that dissipate via Landau damping to determine proton-to-electron heating rates $Q_p/Q_e$.} We distinguish ion-cyclotron frequency circularly polarized waves from low-frequency turbulence and constrain the cascade model using spectra constructed from the latter. We find that the model accurately describes the observed energy spectrum from over 39.4 percent of the intervals from Encounters 1 and 2, \revise{indicating the possibility for Landau damping to heat the young solar wind.} The ability of the model to describe the observed turbulent spectra increases with the ratio of thermal-to-magnetic pressure, $\beta_p$, indicating that the model contains the necessary physics at higher $\beta_p$. We estimate high magnitudes for the Kolmogorov constant which is inversely proportional to the non-linear energy cascade rate. We verify the expected strong dependency of $Q_p/Q_e$ on $\beta_p$ and the consistency of the critical balance assumption.
\end{abstract}

\section{Introduction}

Understanding the mechanisms that drive solar coronal heating and the acceleration of the solar wind are two long-standing problems in space plasma physics. It is crucial to identify the processes that heat protons and electrons to solve these problems. Dissipation of turbulent fluctuations is a likely source for this heating \revise{(\cite{Matthaeus_1999}, see also review in \cite{verscharen2019})}.  Landau damping is one plausible mechanism for the damping of turbulent fluctuations at kinetic scales, whose presence is supported by spacecraft observations \revise{(\cite{Leamon_1999}, \cite{Chen_2019_LD}, \cite{Afshari_2021})}. Observations from the first several encounters of Parker Solar Probe (PSP) \citep{Fox:2015} find that $\beta_p$ is not significantly smaller than unity, implying that Landau damping may be relevant in the young solar wind, defined as the Sun's extended atmosphere below 0.3 au.

Using in situ measurements made by PSP, we model how proton and electron heating rates vary with radial
distance from the Sun, as well as with other plasma conditions. Such determination enables us to better characterize the thermodynamics of this never-before sampled region of the heliosphere, and apply this understanding to other
analogous plasma systems throughout the Universe; that is, those that are hot, diffuse, and weakly collisional (e.g. accretion discs, interplanetary and interstellar medium).

This study addresses the question of turbulent heating by applying a simplified cascade model to PSP observations of the thermal plasma \citep{Kasper2016} and electromagnetic fields
\citep{Bale2016}. The local plasma parameters evaluated using these observations are input into a wavevector-anisotropic steady-state cascade model \citep{Howes2008}. The model assumes a critically-balanced \citep{GS1995, Mallet2015} distribution of low-frequency, Alfv\'enic turbulence from the inertial to dissipation range, \revise{connecting the MHD and kinetic descriptions,} and calculates the linear Landau damping rates as a function of spatial scale perpendicular to the mean magnetic field, producing a steady-state solution for a one-dimensional \cite{Batchelor1953}-like model. \revise{There is increasing evidence from spacecraft observations for the validity of critical balance in the solar wind which is summarized in section 2.2 of \cite{chen_2016}}. The final output of the model is the steady-state energy spectrum of magnetic fluctuations. This spectrum is a function of the spatial scale and the dimensionless Kolomogorov parameters in the model,  $C_1$ and $C_2$, which characterize the rate of non-linear energy cascade and the ratio of the linear-to-nonlinear timescales.
We constrain $C_1$ and $C_2$ using the observed local magnetic field energy spectral density from PSP magnetic field observations. 
The energy of ion-cyclotron frequency, parallel-propagating waves is identified and removed from the observed energy spectrum, retaining only the energy spectral density of low-frequency turbulence. 
The steady-state spectrum evaluated using the model is then used to extract quantities such as spectral indices, the total turbulent heating rate per mass ($Q$), and the relative heating rates for protons ($Q_p$) and electrons ($Q_e$). All of these quantities depend on $\beta_p$, temperature disequilibrium between protons and
electrons ($T_p/T_e$), the strength of the magnetic field ($|\textbf{B}|$), as well as $C_1$ and $C_2$. Here $\beta_p$ is the ratio of the thermal pressure of protons to the magnetic pressure. In this work, we will initially focus on data from the first two
encounters. 

We find that the model describes the local turbulent cascade well for 39.4 $\%$ of the intervals during PSP Encounters 1 and 2, \revise{ indicating that Landau damping is a feasible mechanism for turbulent dissipation in the young solar wind}. The derived heating rates for these intervals are comparable to other empirical estimates \citep{Bandyopadhyay_2020, Hellinger_2013, Martinovic_2020}. The expected strong dependence of the proton-to-electron heating ratio on $\beta_p$ is observed. We estimate the magnitude of the Kolmogorov constant, $C_1$ to be of order 10, which could be due to the inefficiency of energy cascade in solar wind plasma or the shortage in the available energy to cascade due to the imbalance in the flux of Alfv\'enic turbulent fluctuations. We find that the assumption of critical balance in the turbulent cascade is consistent in the young solar wind when the cascade is well described by the model.  

\section{Methodology}
This section provides a brief overview of the procedure employed in this work. A more detailed discussion of the analysis can be found in the appendices. 

\subsection{Numerical method}
\label{ssec:code.cascade}

To numerically determine the steady-state spectrum and associated proton and electron heating rates, we use a code that has been previously applied in
\cite{Howes2008}, \cite{Howes_2011}, and \cite{Kunz:2018}. It assumes a steady-state driving at large scales (small $k_\perp$, where $k_\perp$ is the wavenumber perpendicular to the mean magnetic field), a critically-balanced cascade of energy
through the inertial range, 
\begin{equation}
    \omega = \omega_{\text{non-linear}} = C_2 k_{\perp} v_k,
    \label{eq_intro_1}
\end{equation}

where $\omega$ is the linear frequency of propagation of Alfv\'enic fluctuations, $\omega_{\text{non-linear}}$ is the rate of non-linear energy cascade, $v_k$ is the electron fluid velocity perpendicular to the mean magnetic field, and  $C_2$ is a Kolmogorov-like constant which normalizes the non-linear frequency with respect to the linear frequency. In the dissipation range the code assumes damping onto the protons and electrons described using linear gyrokinetic theory where the damping rate, $\gamma(k_\perp,\mathcal{P})$ (units of [$s^{-1}$]) is a function of $k_\perp$ and a set of plasma parameters, $\mathcal{P}$ that describe the plasma equilibrium \citep{Howes_2006}. For this work, we assume a proton-electron plasma with isotropic temperatures. Therefore $\mathcal{P} = \left[ \beta_p, T_p/T_e \right]$. The code then evolves the following 1-D conservation equation \citep{Batchelor1953} for the magnetic field spectral density until the steady-state is reached,

\begin{equation}
    \dfrac{\partial b_k^2}{\partial t} = -k_{\perp} \dfrac{\partial \epsilon(k_{\perp})}{\partial k_{\perp}} + S(k_{\perp}) -2 \gamma b_k^2 = 0.
    \label{eq_intro_2}
\end{equation}

Here $b_k^2 = \delta B_\perp^2(k_\perp)/4 \pi n_p m_p$ is the square of magnetic field fluctuation amplitude in velocity
units ([$m^2s^{-2}$]), $S(k_{\perp})$ is the source function that inputs energy at turbulence driving scales, and the cascade rate is written as a function of scale as

\begin{equation}
    \epsilon(k_\perp)= C_1^{-3/2} k_\perp \overline{\omega} b_k^3, \hspace{0.3cm} \text{where} \hspace{0.2cm} \overline{\omega}  = \frac{\omega}{k_\parallel v_A}.
    \label{eq_intro_3}
\end{equation}

Here $\epsilon(k_\perp)$ has units of energy per mass per time and $C_1$ is the Kolmogorov constant. The higher the value of $C_1$, the lower the non-linear turbulent energy cascade rate.  We model the turbulent heating rate per unit mass at a given
wavenumber $k_\perp$ as $2 \gamma b_k^2$ (units of [$m^2 s^{-3}$]).

The one-dimensional magnetic energy spectral density is written as $\frac{b_k^2}{k_\perp}$
(units of [$m^3 s^{-2}$]). The code solves equation \ref{eq_intro_2} in dimensionless units over a wide range of scales from the smallest wavenumber, $k_{\perp i} \rho_p = 0.01$  to the largest wavenumber, $k_{\perp f} \rho_e = 1.2$. Here $\rho_p$ and $\rho_e$ are the proton and the electron gyroradii respectively.

The  proton and electron
heating rates are given by 
\begin{equation}
  Q_s = \int dk_\perp 2 C_2 \left( \frac{\gamma_s}{\omega}\right) \overline{\omega} b_k^3.
\end{equation}

The magnetic field spectral density  and heating rates are normalized to physical units via $b_{ki}$,  the amplitude of magnetic field fluctuations at the largest scale considered in the code, $k_{\perp i}$. A detailed discussion of the normalization of code outputs to physical units can be found in appendices \ref{app_model_spectrum} and \ref{app_normalization}.

\subsection{In situ data sets}
The PSP/FIELDS instrument suite makes in situ observations of electromagnetic fields (\cite{Bale2016}). We use level 2 magnetic field data (in RTN coordinates) from the flux gate magnetometer (MAG). The PSP/SWEAP instrument suite makes in situ observations of solar wind thermal plasma that are processed to evaluate the velocity distribution functions (VDFs) of ions and electrons and associated moments (\cite{Kasper2016},\cite{Case_2020}
\cite{Whittlesey_2020}, \cite{Livi_2021}). For this study, we use proton VDF fits from SPC and electron VDF fits from SPANe \citep{Halekas2020}. The FIELDS and SWEAP observations are prescreened for ``good" intervals of length $\sim$ 15 minutes as elaborated in Appendix \ref{app_sec_data} , obtaining 1072 and 1046 intervals in Encounters 1 and 2 respectively. \revise{We choose 15 minute intervals so that the lowest resolved spectral frequencies are in the inertial range of the energy spectrum. This allows us to model the damping that arises over the transition between the inertial and dissipation ranges. Furthermore, this choice increases the statistical leverage of our analysis.} Switchback filtration is not done on the observations as \cite{Martinovic_2021} have shown that the observations are fairly similar inside and outside of switchbacks. Average values of the plasma parameters over the 15 minute intervals are used as inputs for the cascade model.

PSP has reaction wheels to maintain the Sun-facing orientation of the spacecraft during encounters. These wheels generate coherent, sharp-peaked noise in the energy spectrum at the rotation frequencies of the wheels, as well as harmonic and beat frequencies.  For each interval, the reaction wheel noise is identified and removed, as described in Appendix \ref{app_noise_removal}.

 During Encounter 1, the sampling rate of FIELDS/ MAG varies between  $\sim$ 73.24 Sa/s, 146.48 Sa/s, and 292.96 Sa/s. During Encounter 2, the magnetic field is sampled at a constant rate of 146.48 Sa/s. We consider the energy spectrum up to 10 Hz as the energy spectrum hits the noise floor at $\sim$ 10 Hz for Encounters 1 and 2 \citep{Bowen_noise}. A Butterworth low-pass filter of order 10 and a cut-off frequency of 18.31 Hz is applied  and the magnetic field observations are downsampled to a sampling rate of $\sim$ 36.62 Sa/s. More details on the downsampling approach are found in Appendix \ref{app_downsampling}.

\subsection{Evaluation of energy spectral density of turbulent fluctuations using PSP observations }
 A Morlet wavelet transform is employed to evaluate the energy spectral density of turbulent magnetic field fluctuations. Wavelet transforms resolve the energy of a signal in both time and frequency (See \cite{Terrence_compo} and \cite{Podesta_2009} for introductory details on these transforms).

The wavelet energy spectrum of the reaction wheel noise-removed, anti-aliased magnetic field time series, $|\tilde{B}|^2_{psp}$,  is evaluated. More details on the evaluation of the wavelet energy spectral density are found in Appendix \ref{app_wavelet_psp}.

Parallel propagating Ion Cyclotron Waves (ICWs) and Fast Magnetosonic/Whistler waves (FM) are often identified near kinetic scales in Encounters 1 and 2 observations (\cite{Bowen_ICW}, \cite{Verniero_2020})
The \cite{Howes2008} model assumes a cascade of low-frequency Alfv\'enic turbulent fluctuations and doesn't account for the presence of these ion-cyclotron frequency waves. 
For each interval, the energy due to the ion-cyclotron frequency coherent waves, $|\tilde{B}|^2_{wave\_psp} (k_\perp \rho_p)$, is identified and removed from the observed energy spectrum, $|\tilde{B}(f)|_{psp}^2$, retaining the low-frequency turbulent energy spectrum, $|\tilde{B}|^2_{turb\_psp}(k_\perp \rho_p)$, as described in Appendix \ref{app_high_f_wave_removal}. Taylor's hypothesis \citep{Taylor1938} is then used to transform the spectrum from depending on frequency to spatial scales. \revise{\cite{Perez_2021}} have shown that Taylor's hypothesis is a fair assumption to be made for the observations from the early encounters of PSP.

\subsection{Constraining $C_1$, $C_2$ }

For each interval, the model energy spectrum, $|\tilde{B}|^2_{model} (k_\perp \rho_p)$   is evaluated by employing the \cite{Howes2008} cascade model. The Kolmogorov constants $C_1$ and $C_2$ in the model are constrained by minimizing the difference between the two energy spectra, $|\tilde{B}|^2_{turb\_psp}$ and  $|\tilde{B}|^2_{model}$. The function, $R^2 (C_1,C_2)$ is minimized with respect to $C_1$ and $C_2$, where, 
\begin{equation} \label{Eq_R2}
\begin{split}
    & R^2(C_1,C_2) = \\
    & \sum_{j \in k_\perp} \dfrac{\left[ \log \left(|\tilde{B}|^2_{turb\_psp} (k_{\perp,j} \rho_p) \right) - \log \left(|\tilde{B}|^2_{model} (k_{\perp,j} \rho_p,C_1,C_2) \right) \right]^2}{\log \left(|\tilde{B}|^2_{turb\_psp}(k_{\perp,j} \rho_p) \right)^2}.
    \end{split}
\end{equation}
$R^2$ is initially evaluated over a dense grid of ($C_1, C_2$) values and a set of local minima is identified. All the local minima are refined using the Levenberg-Marquardt algorithm. The refined local minimum corresponding to the least value of $R^2$ is considered to be the absolute minimum. A more detailed description of constraining  $C_1$ and $C_2$ is found in Appendix \ref{app_C1_C2_constrain}.

The magnitude of the turbulent spectrum varies over a wide range from inertial to dissipation scales. The magnitude of $R^2$ is more sensitive to the inertial scales compared to the dissipation scales. In order to better quantify the difference between $|\tilde{B}|^2_{turb\_psp}$ and  $|\tilde{B}|^2_{model}$ in dissipation scales, we define a second goodness-of-fit metric that focuses specifically on the scales where the proton and electron damping occurs, 

\begin{equation}
\frac{\Delta E}{E}_{diss} = \frac{\sum_{(k_\perp \rho_p)_{diss}} | |\tilde{B}|^2_{turb\_psp} - |\tilde{B}|^2_{model} | \Delta (k_\perp \rho_p) }{\sum_{(k_\perp \rho_p)_{diss}} |\tilde{B}|^2_{turb\_psp} \Delta (k_\perp \rho_p) }.
\end{equation}
Here we identify the dissipation region, $(k_\perp \rho_p)_{diss}$ as the region between the scale where the spectral index of $|\tilde{B}|^2_{model}$ steepens to -2.5 and the scale corresponding to the highest frequency considered (10 Hz). The spectral index threshold of -2.5 is empirically chosen so that $(k_\perp \rho_p)_{diss}$ begins at scales where the slope of the model energy spectra around the spectral break has reached its dissipation range value. Variations in the exact value of this threshold do not qualitatively change the number of intervals well described by the cascade model. This selection allows the deviation of  $|\tilde{B}|^2_{turb\_psp}$ from  $|\tilde{B}|^2_{model}$ at dissipation scales to be better classified.
In intervals where the best-fit $R^2 \le 0.03$ and $\frac{\Delta E}{E}_{diss} \le 0.25$, the turbulent cascade is considered to be well described by the model. In such intervals, heating rates, $Q_p$, $Q_e$, and $Q_{total}$ are evaluated using best-fit values of Kolmogorov constants as described in Appendix \ref{app_normalization}. 

Figure \ref{fig_project_overview} illustrates the overview of the methodology of this work for four intervals: (a) An interval where the turbulent cascade is well described by the model and ion-cyclotron frequency coherent waves are not observed, (b) An interval where an ion-cyclotron frequency coherent wave is observed, its energy is identified and removed and the turbulent cascade is well described by the model, (c) and (d) two intervals where the turbulent cascade is not accurately described by the model, with (c) the best-fit $R^2 > 0.03 $ and (d) the best-fit $R^2 < 0.03 $.  
\begin{figure*}
    \centering
    \plotone{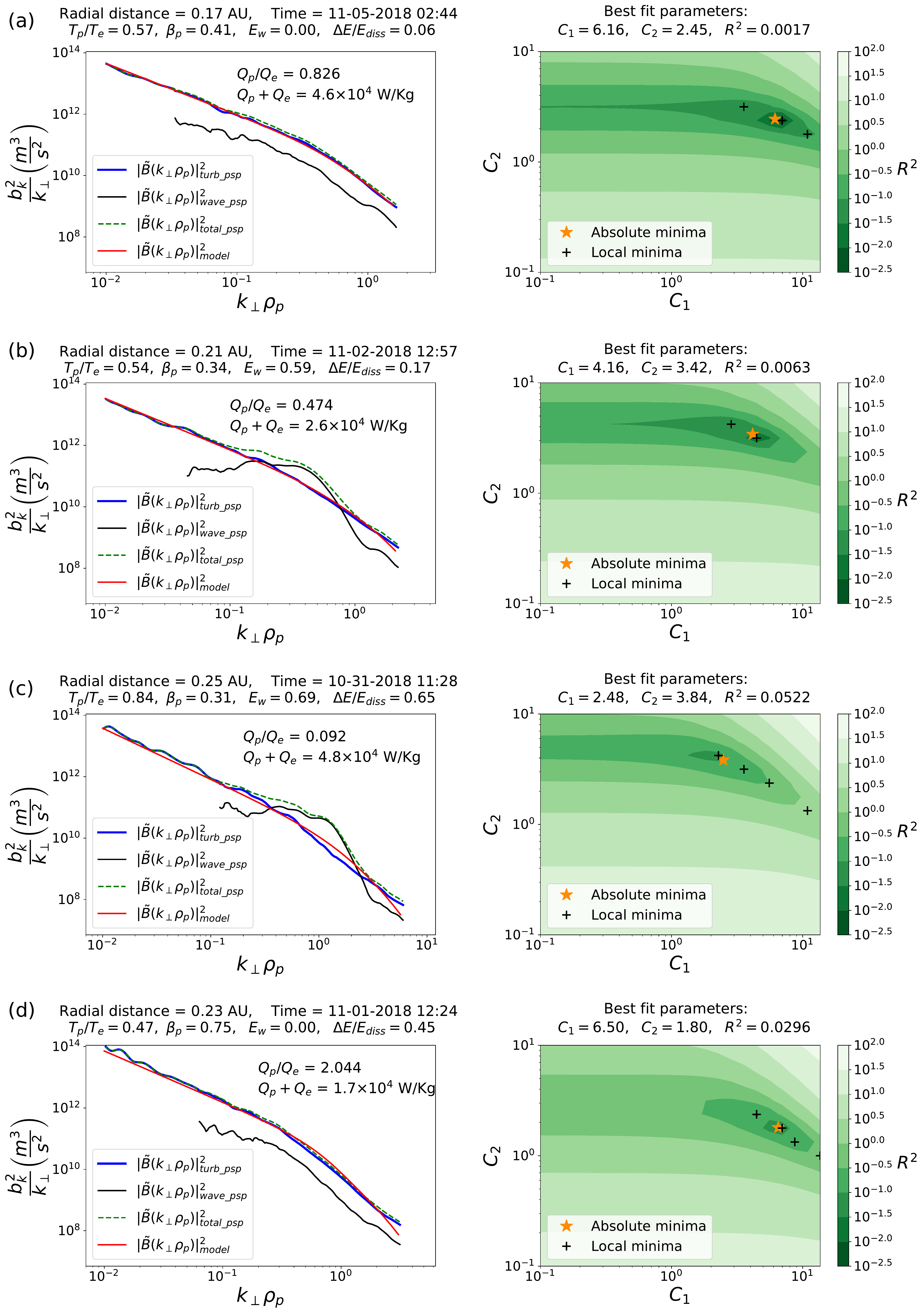} 
    \caption{Overview of the methodology for four example intervals a,b,c, and d. The right panel shows the contour map of $R^2$ as a function of the initial ($C_1, C_2$) grid values, the local minima (black pluses), and the absolute minima (orange star). The left panel shows the total observed energy spectrum (green dashed), the observed turbulent energy spectrum (blue), the observed energy due to ion-cyclotron frequency waves (black), the best-fit model energy spectrum (red) corresponding to the absolute minima, and the evaluated proton and electron heating rates. Here $E_w$ is a parameter that quantifies the fraction of narrow-band, ion-cyclotron frequency coherent wave energy. The physical differences between the four intervals are described in the text.}
    \label{fig_project_overview}
\end{figure*}


\section{Relevance of Landau damping in the young solar wind}
\label{sec_rel_LD}

\begin{figure*}[!h]
         \centering
            \plotone{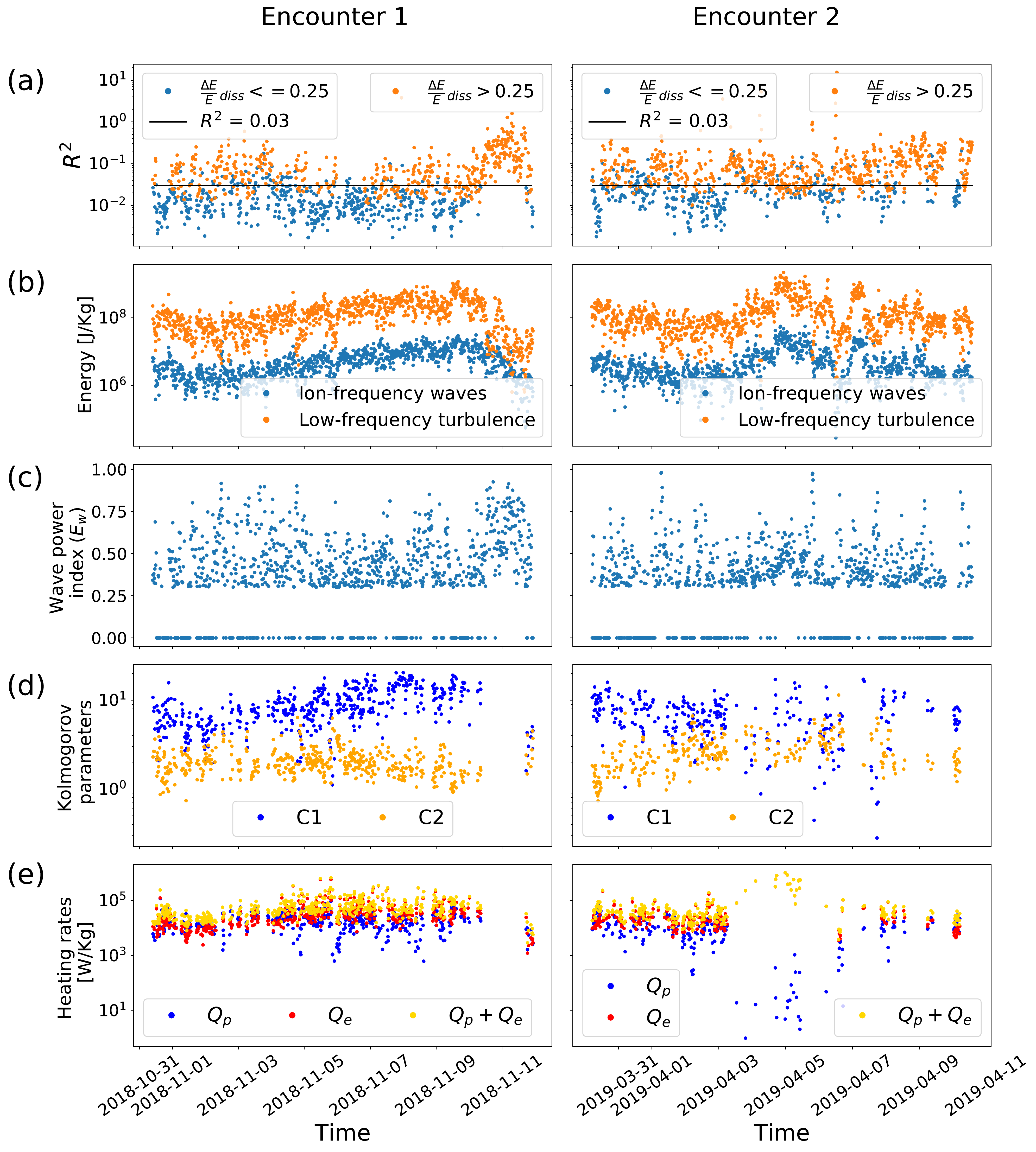} 
            \caption{Cascade model properties extracted from first two PSP encounters as a function of time: (a) $R^2$ profile where $\frac{\Delta E}{E}_{diss} \le 0.25$ (blue), $\frac{\Delta E}{E}_{diss} > 0.25$ (orange), $R^2$ threshold (black) of 0.03, (b) energy in ion-cyclotron frequency coherent waves (blue) and turbulent fluctuations (orange), and (c) $E_w$ for all intervals. (d) Best fit Kolmogorov parameters, $C_1$ (blue), $C_2$ (orange), (e) proton ($Q_p$, blue), electron ($Q_e$, red) and total ($Q_{total}$, yellow) heating rates with respect to time for intervals well described by the \cite{Howes2008} model.}
            \label{fig_LD_E1_E2_relevance}
\end{figure*}

Using the criteria discussed in section 2, the \cite{Howes2008} cascade model, which dissipates turbulent energy entirely via proton and electron Landau damping, describes the turbulent cascade accurately in 39.4 $\%$ of intervals in Encounters 1 and 2. \revise{That this model fits
the observations is consistent with the value of $\beta_p$ not being much smaller than unity for these encounters where the median value is 0.27 (panel (a) of Figure \ref{fig_model_beta_eff})}. The main results of this work have been summarized in Figure \ref{fig_LD_E1_E2_relevance}. Panel (a) shows the profile of best-fit $R^2$ with respect to time for all intervals from the first two PSP encounters. The goodness-of-fit for the dissipation scales is indicated by color, with $\frac{\Delta E}{E}_{diss} \le 0.25 (>0.25)$ shown with blue (orange) dots. Panel (b) shows the energy in ion-cyclotron frequency waves and low-frequency turbulence for all intervals, calculated by integrating the spectra, $|\tilde{B}|^2_{wave\_psp}$ and $|\tilde{B}|^2_{turb\_psp}$ respectively.

\begin{figure}[!h]
    \centering
    \plotone{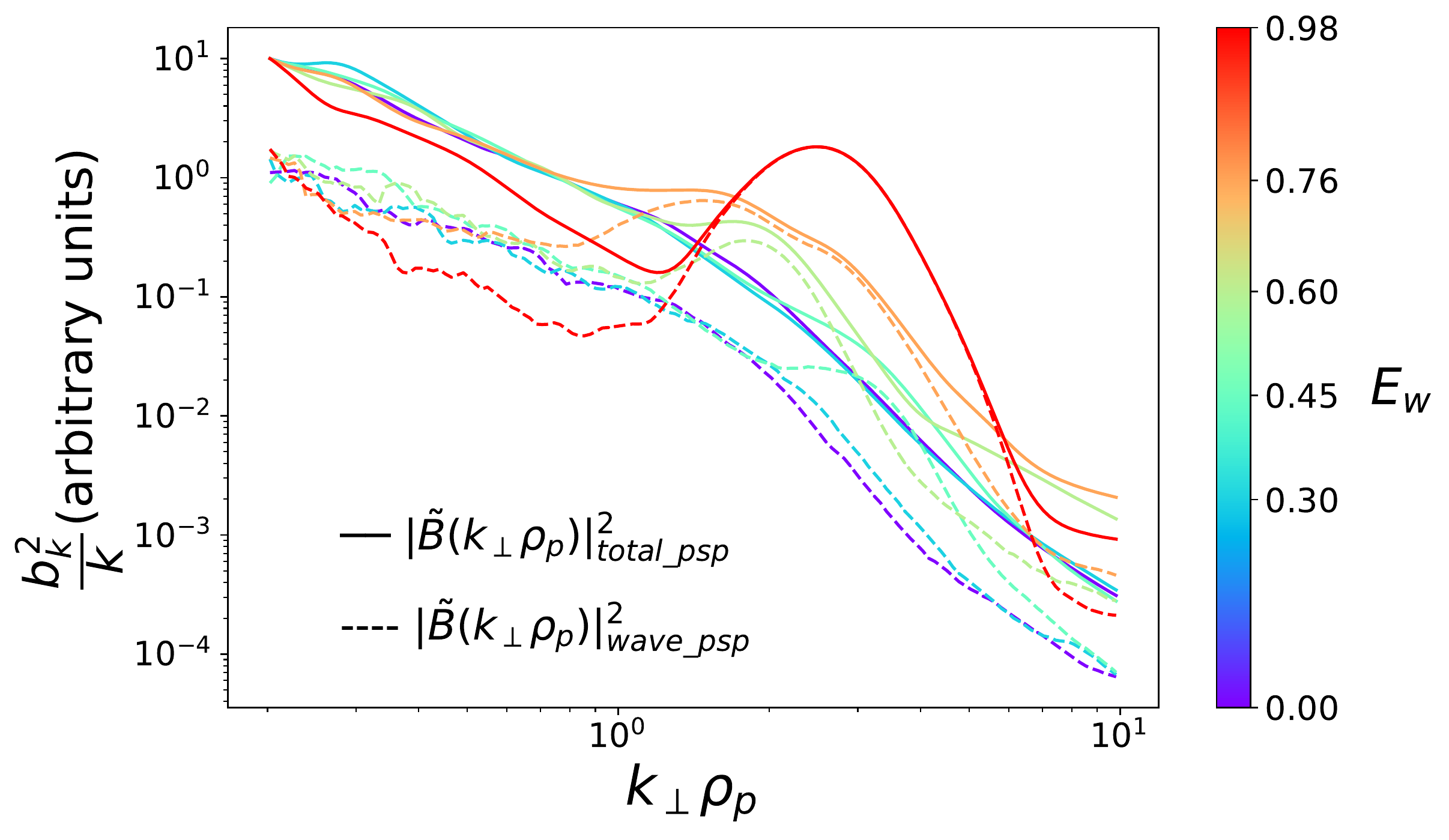} 
    \caption{Energy spectra for 6 example intervals with  ion-cyclotron frequency coherent waves and their corresponding values of $E_w$ on the colorbar. The energy spectra have been normalized such that the total energy (turbulent + ion-cyclotron frequency waves) at 0.2 Hz is 10 for all intervals.  }
    \label{fig_Pw_compare}
\end{figure}

Furthermore, we define a parameter $E_w$ (Figure \ref{fig_Pw_compare}) to characterize the fraction of the narrow-band, ion-cyclotron frequency coherent wave energy in an interval. The values of $E_w$ lie between 0 and 1. The ion-cyclotron frequency waves are observed at frequencies greater than 0.2 Hz in the first two PSP encounters. Our energy spectra are evaluated at $f_{log\_bin}$, the median values of logarithmic spaced frequency bins, $\Delta f_{log\_bin}$ (discussed in Appendix \ref{app_wavelet_psp}). Random coherences between perpendicular wavelet transform components can result in up to 30 percent of power at each frequency bin being erroneously attributed to waves (Identifying energy of ion-cyclotron frequency waves is discussed in Appendix \ref{subsec_cross_coherence}). Defining the set of frequencies in $f_{log\_bin}$ which are greater than 0.2 Hz and at which the ratio of the coherent wave energy to total energy is greater than 0.3 as $f_{wave}$, $E_w$ is calculated by evaluating the ratio of summed wave energy over $f_{wave}$ to the summed total energy over $f_{wave}$. 

\begin{equation}
\begin{gathered}
    E_{w} = \dfrac{\sum_{f_{wave}} |\tilde{B}|^2_{wave\_psp}(f_{log\_bin}) \Delta f_{log\_bin}}{\sum_{f_{wave}} |\tilde{B}|^2_{psp} (f_{log\_bin}) \Delta f_{log\_bin}} , \\ \text{where}  \hspace{0.4cm}   f_{wave} = \left\{ f_{log\_bin} \middle| f_{log\_bin} > 0.2 \text{Hz} \hspace{0.4cm} \cup \right. \\ \left. \dfrac{|\tilde{B}|^2_{wave\_psp}(f_{log\_bin})}{|\tilde{B}|^2_{psp}(f_{log\_bin})} \ge 0.3 \right\} .
\end{gathered}
\end{equation}

 The parameter $E_w$ is defined so as to consider the energy in only those frequencies where the ion-cyclotron frequency waves occur. Figure \ref{fig_Pw_compare} shows six PSP intervals with ion-cyclotron frequency waves occurring over varying ranges of frequencies and with narrow but varying bandwidths. The energy spectrum in purple has no coherent waves and the corresponding value of $E_w$ is 0. The fraction of energy in coherent waves increases along the colorbar and the value of $E_w$ increases accordingly. The energy spectrum in red has one of the highest fractions of energy in waves among all intervals and the corresponding value of $E_w$ is 0.98. Panel (c) in Figure \ref{fig_LD_E1_E2_relevance} shows the profile of $E_w$ for all intervals. Forty one percent of intervals in the first two PSP encounters have $E_w \ge 0.4 $.

Panel (d) shows the best-fit values of $C_1$ and $C_2$ for intervals well described by the model. In general, the magnitude of $C_2$ is of order unity and is consistent with our assumption of critical balance. However, the magnitude of $C_1$ we estimate is of order 10 in contrast to neutral fluids, where its value is $\sim$ 0.5 \citep{KC1_neutral}. This overestimation is partly due to the inefficiency of turbulent cascade in plasmas compared to neutral fluids due to the breaking of strong magnetic field lines occurring in the former. \cite{Beresnyak_2011} measured the value of $C_1$ to be $\sim$ 4.2 in MHD turbulence simulations. Another reason for the overestimation of $C_1$ may be the shortage of actual energy available to be cascaded due to the imbalance in turbulent energy flux. Cross helicity, $\sigma_c$, quantizes this imbalance and is evaluated using the time series of magnetic field vector, \textbf{B}(t) and solar wind velocity vector, $\textbf{V}_{sw}$ at an inertial range frequency of $8.3 \times 10^{-3}$ Hz (2 minutes) as described in equations 3-10 in  \cite{Wicks_2013} for all intervals and is then averaged over each interval. A $\sigma_c$ value of: +(-)1 implies a pure anti-sunward (sunward) directed turbulent flux and a value of 0 implies a balanced turbulent flux. \revise{The distribution of the interval-averaged $\sigma_c$ values indicates a highly imbalanced anti-sunward directed turbulent flux in Encounters 1 and 2. The lower quartile, median and upper quartile values for this distribution are 0.75, 0.85, and 0.92 respectively. The corresponding values of percentage reduction in cascade rate (evaluated using equation 6 in \cite{Matthaeus_2004}) are 40, 54, and 67 respectively.} The model assumes that the cascade is balanced and all the energy in the magnetic field fluctuations at the largest scale, expressed by equation \ref{norm9}, is available to be cascaded. However, in reality, just a small fraction of this energy could be available to cascade due to the imbalance, leading to an overestimation of $C_1$. The magnitude of the Pearson correlation coefficient between $C_1$ and cross-helicity is 0.47, which supports this interpretation. Additional statistical comparisons can be found in section \ref{sec_qp_qe_parameters}.


\begin{figure*}
    \centering

    \plotone{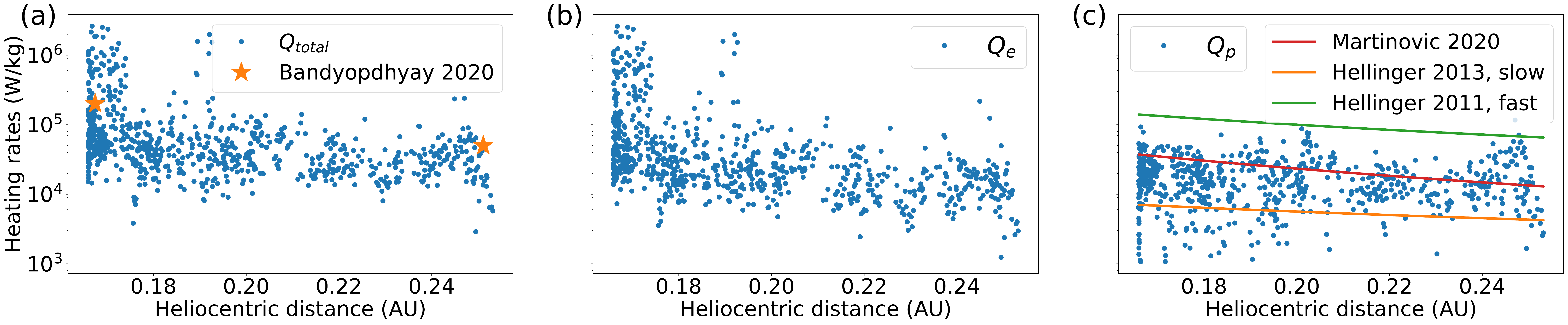} 
    \caption{ Radial profiles of heating rates: (a) total heating rate estimated in this work, $Q_{total}$ (blue) and the estimation by \cite{Bandyopadhyay_2020} at 36 and 54 solar radii (orange stars), (b) electron heating rate estimated in this work, $Q_e$ (blue), (c) proton heating rate estimated in this work, $Q_p$ (blue), extrapolation of the estimation of proton heating rates by \cite{Hellinger_2013} in the slow solar wind(orange dashed) and \cite{Hellinger_2011} in the fast solar wind (green dashed) using Helios observations and estimation of proton heating via stochastic heating by \cite{Martinovic_2020} (red dashed).  }
    \label{fig_Heating_rates}
\end{figure*}

Panel (e) in Figure \ref{fig_LD_E1_E2_relevance} shows the estimated profiles of the electron, proton, and total heating rates due to Landau damping in intervals that are well described by the model. Figure \ref{fig_Heating_rates} shows these heating rates as a function of heliocentric distance along with several empirical estimates. \cite{Bandyopadhyay_2020} estimated a fluid scale energy transfer rate at 36 and 54 solar radii. \cite{Martinovic_2020} estimated proton heating rates due to stochastic heating using the amplitude of turbulent velocity fluctuations near the ion gyroscales. \cite{Hellinger_2013} estimated proton heating rates in the slow wind between 0.3 and 1 AU using radial fits of magnetic field strength, proton number density, parallel and perpendicular temperatures, and heat fluxes evaluated from Helios 1 and 2 observations. The right panel of Figure \ref{fig_Heating_rates} shows the extrapolation of their estimation down to 0.16 AU. Our estimations are comparable to all of the above distinct empirical estimates. That our calculations are comparable to other methods provides support for these estimations of the heating rates.

\subsection{Statistical comparison between best-fit $R^2$ and parameters}

To determine what plasma and solar wind parameters describing the underlying turbulence influence the cascade model, we perform a statistical comparison between the parameters and the goodness of fit values (best-fit $R^2$ and $\frac{\Delta E}{E}_{diss}$). The statistics with either of the goodness of fit values are qualitatively similar and thus we report only for best-fit $R^2$. The Coulomb number is written in terms of proton-proton collision frequency, $\nu_{pp}$ and solar wind speed, $V_{sw}$ as 
\begin{equation}
    N_c =  \frac{\nu_{pp} R}{V_{sw}}.
\end{equation}
Here $\nu_{pp}$ is evaluated as described in equation 2b in Appendix B of \cite{Wilson_III_2018}. We observe that the energy spectra of many intervals in the first two PSP encounters fall steeper than $k_\perp^{-2.8}$ (a value expected for balanced KAW turbulence \citep{Howes_2011_PRL} )  over a ``transition" region (\cite{Bowen_spectral_steepening} and \cite{Squire_2022}) at the beginning of the dissipation range. We define $\alpha_{max}$ to quantify this steepness of the slope. The dissipation range in PSP Encounters 1 and 2 starts at a frequency greater than 0.2 Hz. We consider $f_{log\_bin}$ greater than 0.2 Hz and group it in equal width-bins of $\sim$ 40 data points. In every bin a linear fit of log($|\tilde{B}|^2_{turb\_psp}$) vs log($f_{log\_bin}$) is performed and the slope is determined. $\alpha_{max}$ is defined as the steepest i.e. the minimum of these slopes.

\begin{figure}
    \centering
    \plotone{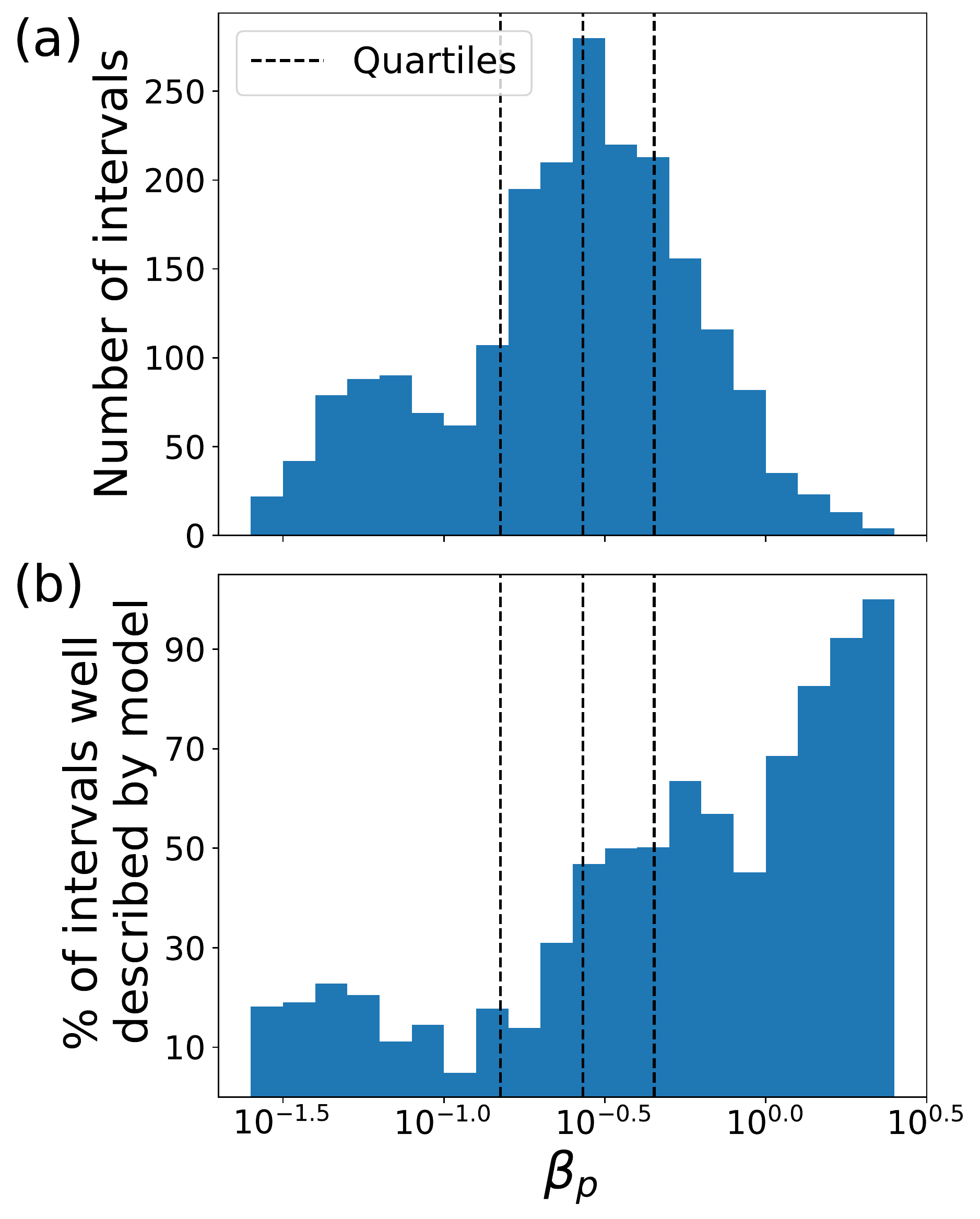}
    \caption{\revise{(a) Distribution of intervals obtained from Encounters 1 and 2, and (b) the percentage of those intervals} well described by \cite{Howes2008} model as a function of $\beta_p$. The accuracy of the model in describing the observed turbulent energy spectrum roughly increases with $\beta_p$ implying that Landau damping could be the dominant process responsible for heating the young solar wind in higher $\beta_p$ regions.}
    \label{fig_model_beta_eff}
\end{figure}

The Pearson correlation coefficient ($r_p$) quantizes the linearity between two parameters while the Spearman correlation coefficient ($r_s$) quantizes the monotonicity between two parameters. Table \ref{R2_correlation} contains both these correlation coefficients between best-fit $R^2$ and the following interval-averaged parameters: $\beta_p$, square of magnetic field amplitude ($|\textbf{B}|^2$), $T_p/T_e$, $\sigma_c$, $V_{sw}$, heliocentric distance (R), and $N_c$ as well as $E_w$ and $\alpha_{max}$.  From Table \ref{R2_correlation} we can infer that best-fit $R^2$ has a substantial non-linear correlation with the parameters $\beta_p$, $T_p/T_e$, $\alpha_{max}$ and $E_w$. The strong correlation with $E_w$ is partly due to the inefficiency of our criteria in separating the turbulent energy spectrum as a smooth curve from the ion-frequency wave energy spectrum when large amplitude waves are present. The high correlation with $\alpha_{max}$ is expected as the model doesn't account for a steep transition region. The notable correlation with $T_p/T_e$ is because of the correlation between $T_p/T_e$ and $\beta_p$. The significant negative correlation with $\beta_p$ is because the model doesn't encompass all the necessary physics to describe the dissipation of turbulent fluctuations in the low $\beta_p$ regime. \revise{In Encounters 1 and 2, the ability of the model to accurately describe the observed turbulent power spectra roughly increases with $\beta_p$. The model accurately describes the observed spectra in $16.8, 27.2, 52.0$ and $61.6\%$ of the intervals for the first through fourth quartiles as organized by $\beta_p$ as illustrated in Figure \ref{fig_model_beta_eff}.}

\begin{table}[!h]
    \setlength{\tabcolsep}{2pt}
    \begin{tabular}{ |c|c|c|c|c|c|c|c|c|c| } 
 \hline
$X$ $\longrightarrow$ & $\beta_p$  & $|\textbf{B}|^2$ & $T_p/T_e$  & $E_w$ & $\alpha_{max}$ & $\sigma_c$ & $V_{sw}$ & R &  $N_c$ \\ [0.5ex] 
\hline
$r_{p} \left(R^2,X\right)$  & -0.08 & 0.02 & -0.09 & 0.13 & 0.08 & -0.06 & -0.07 & 0.03 &  0.06\\ 
\hline
$r_{s} \left(R^2,X\right)$  & -0.41 & -0.05 & -0.35 & 0.52 & 0.44 & 0.08 & -0.24 & 0.22  & 0.28\\ 
 \hline
\end{tabular}
    \caption{Pearson and Spearman correlation coefficients of best-fit $R^2$ with solar wind plasma parameters for all intervals from Encounters 1 and 2.}
    \label{R2_correlation}
\end{table}

\section{Behavior of $Q_p/Q_e$ with respect to plasma parameters} 
\label{sec_qp_qe_parameters}

\begin{figure*}[!h]
    \centering
    \plotone{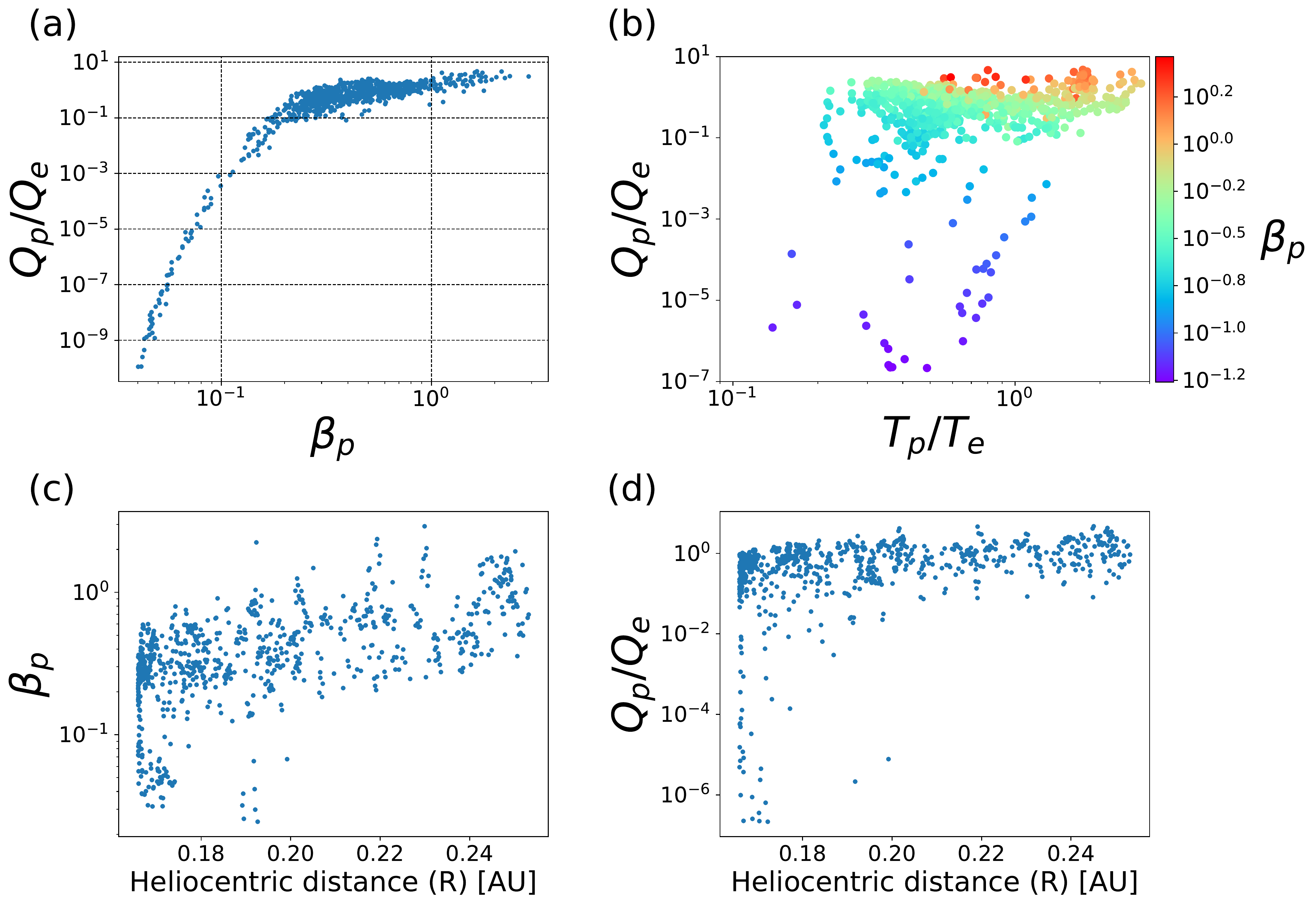} 
    \caption{Behaviour of $Q_p/Q_e$ with plasma parameters: The profiles of $Q_p/Q_e$ with respect to (a) $\beta_p$, (b) $T_p/T_e$, the profiles of (c) $\beta_p$, and (d) $Q_p/Q_e$ with respect to heliocentric distance for intervals where the cascade is well described by the model.}
    \label{fig_qp_qe_vs_plasma_params}
\end{figure*}

To determine what parameters influence the distribution of heating between protons and electrons, we evaluate both the Pearson and Spearman correlation coefficients of the ratio of proton-to-electron heating rate, $Q_p/Q_e$ and various parameters discussed in section \ref{sec_rel_LD} (Table \ref{Qp_qe_correlation}). The estimated magnitude of $Q_p/Q_e$ has a high non-linear correlation with $\beta_p$ which subsumes the expected anti-correlation with $|\textbf{B}|^2$. A similar correlation between $|\textbf{B}|^2$ and proton temperature has been measured from ACE observations in \cite{Smith_B2_T_p_correlation}. The large correlation with heliocentric distance, R is due to the high correlation between R and $\beta_p$ ($r_p$ = 0.45). We infer that $Q_p/Q_e$ is influenced significantly by $\beta_p$ alone. Figure \ref{fig_qp_qe_vs_plasma_params} reflects the same. 
   
\begin{table}[!h]
    \centering
    \setlength{\tabcolsep}{2pt}
    \renewcommand{\arraystretch}{1.5}
    \begin{tabular}{ |c|c|c|c|c|c|c|c|c|c| } 
 \hline
 $X$ $\longrightarrow$ & $\beta_p$  & $|\textbf{B}|^2$ & $T_p/T_e$  & $\alpha_{max}$ & $\sigma_c$ & $V_{sw}$ & R &  $N_c$ \\ [0.5ex] 
\hline
$r_{p} \left(\frac{Q_p}{Q_e},X\right)$  & 0.79 & -0.63 & 0.3 & -0.10 & 0.02 & 0.11 & 0.49 & -0.13 \\ 
\hline
$r_{s} \left(\frac{Q_p}{Q_e},X\right)$  & 0.82 & -0.69 & 0.24 & -0.10 & 0.06 & 0.05 & 0.48 & -0.18 \\ 
 \hline
\end{tabular}
    \caption{Pearson and Spearman correlation coefficients of $Q_p/Q_e$ with solar wind plasma parameters for intervals with a good fit to the cascade model in Encounters 1 and 2.}
    \label{Qp_qe_correlation}
\end{table}

\cite{Howes_2011} prescribed an analytical expression for $Q_p/Q_e$ as a function of $T_p/T_e$ and $\beta_p$.
\begin{equation}
    \begin{gathered}
        (Q_p/Q_e)_{Howes} = a_1 \frac{a_2^2 + \beta_p^{\alpha}}{a_3^2 + \beta_p^{\alpha}} \sqrt{\frac{m_p T_p}{m_e T_e}} e^{-1/\beta_p}, \\
         \text{where}, \hspace{0.5 cm} a_1 = 0.92, \hspace{0.5cm} \alpha = 2 - 0.2 \log(T_p/T_e) \\
         a_2 = 1.6/(T_p/T_e), \hspace{0.5cm} a_3 = 18 + 5\log(T_p/T_e)
    \end{gathered}
\end{equation}

\begin{figure}[!h]
    \centering
    \plotone{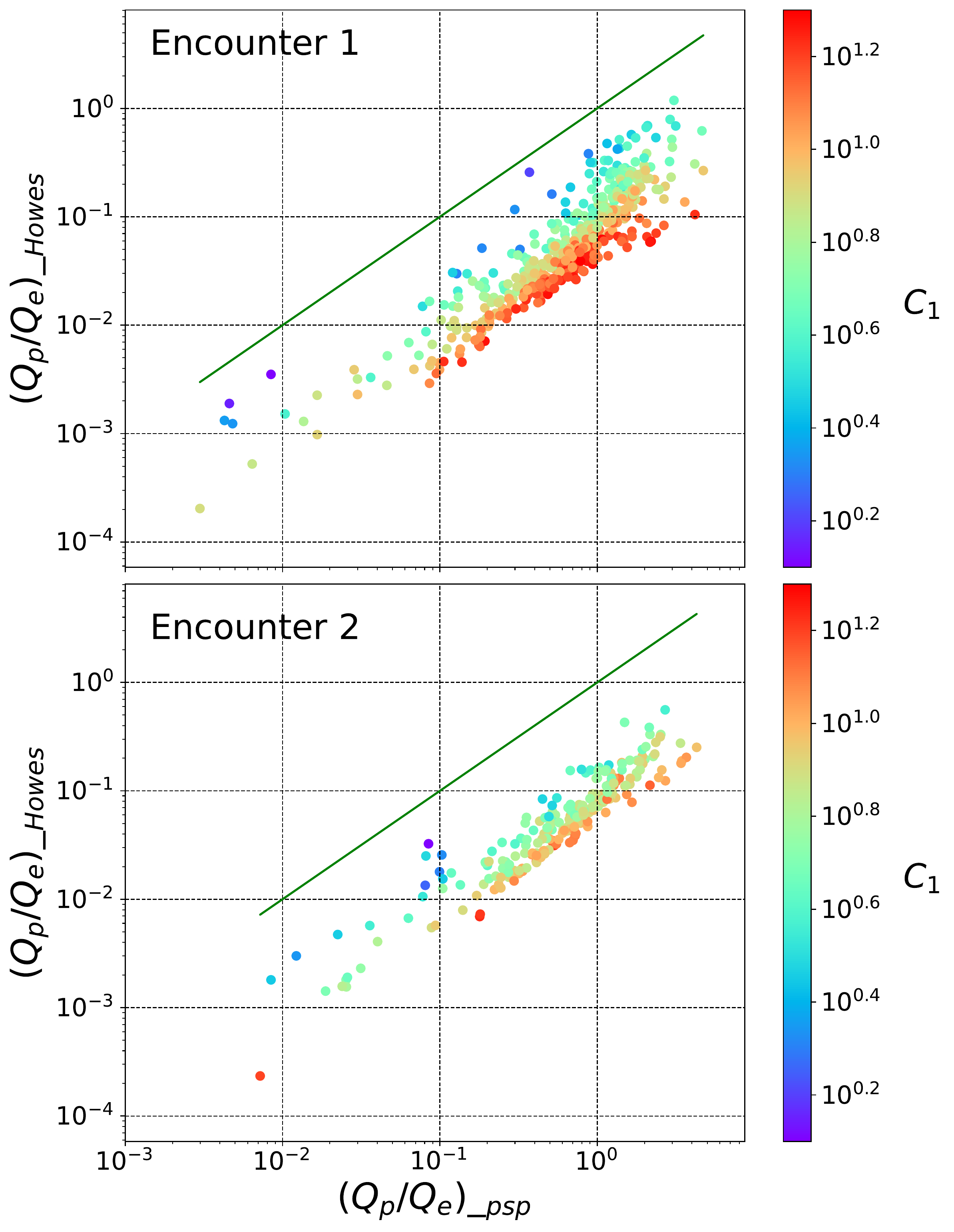} 
    \caption{Comparison of $Q_p/Q_e$ prescribed by \cite{Howes_2011} and $Q_p/Q_e$ estimated using PSP observations in this work along with the dependence on $C_1$ (colorbar). The higher our estimated value of $C_1$, the greater our estimate of $Q_p/Q_e$ as compared to \cite{Howes_2011}.  }
    \label{fig:QiQe_Howes_comparision}
\end{figure}

Figure \ref{fig:QiQe_Howes_comparision} shows the comparison of our estimates of $Q_p/Q_e$ with the \cite{Howes_2011} prescription. In general, we overestimate the heating ratio compared to the prescription which is due to the significantly larger values of C1 that are extracted from our fits compared to the values used for the prescription. The higher the magnitude of $C_1$, the slower the non-linear cascade, and more time is available for the linear damping processes to heat the protons before the cascade reaches the electron scales.  

\section{Discussion}
We find that the low-frequency, Alfv\'enic cascade model describes the observed magnetic spectrum well for 39.4 $\%$ of the intervals during PSP Encounters 1 and 2, \revise{indicating that Landau damping is a feasible mechanism for turbulent dissipation in the young solar wind}. The ability of the model to describe the observed
\revise{energy spectra increases with $\beta_p$, increasing from $16.8 \%$ for $\beta_p < 0.15$ to $61.6\%$ for $\beta_p> 0.45$}. We estimate the magnitude of the Kolmogorov constant, $C_1$ to be of order 10. The high magnitude of  $C_1$ could be due to the inefficiency of the energy cascade in solar wind plasma or the shortage in the available energy to cascade due to the imbalance in the turbulent energy flux. The expected strong dependence of the ratio of proton-to-electron Landau damping on $\beta_p$ is verified.  The assumption of a critically balanced turbulent cascade is found to be consistent in the young solar wind when the spectrum is well described by the model. We verify that the heating rates we estimate are comparable to other empirical estimates. 

One of the main assumptions that the \cite{Howes2008} model makes, contrary to the observations, is that the turbulent cascade is balanced. This assumption is partly responsible for high $C_1$ estimations and requires the qualification of any comparison of $Q_p/Q_e$ from balanced models of turbulence with observations. We intend to account for the imbalance in turbulent flux in future work. \revise{Furthermore, the assumption of critical balance and low-frequency fluctuations made with this model may not always hold. Spacecraft observations have been described using other models of turbulence e.g. \cite{CB_Adhikary_2022} and \cite{CB_Telloni_2022}. While observed ion-cyclotron frequency fluctuations are identified and removed in this work, the contribution of their damping will be considered in future work.} Additionally, we ensure that the observed energy spectrum agrees with the model energy spectrum only over ion scales, up to frequencies where the measurements encounter the noise floor. There are no spectral constraints at higher frequencies, where electron processes are acting. 



\section{Acknowledgements}
KGK and NS were supported by NASA Grants 80NSSC19K0912 and 80NSSC20K0521. MMM and KGK were supported by NASA grant 80NSSC22K1011. 
The authors thank G. Howes for providing the original code for the cascade model used in \cite{Howes2008}. Parker Solar Probe was designed, built, and is now operated by the Johns Hopkins Applied Physics Laboratory as part of NASA's Living with a Star (LWS) program (contract NNN06AA01C). Support from the LWS management and technical team has played a critical role in the success of the Parker Solar Probe mission. This work was performed in part at Aspen Center for Physics, which is supported by National Science Foundation grant PHY-1607611. PSP/SWEAP and FIELDS data can be accessed at: \url{https://hpde.io/NASA/NumericalData/ParkerSolarProbe/SWEAP/index.html} and \url{https://hpde.io/NASA/NumericalData/ParkerSolarProbe/FIELDS/index.html}.
 
\appendix

\section{Data} \label{app_sec_data}

For each interval, the energy spectral density of the turbulent magnetic field fluctuations, $|\tilde{B}(f)|_{turb\_psp}^2$, is evaluated from PSP observations using the following steps:
\begin{enumerate}

\item \textbf{Section \ref{app_sec_data}:} Pre-screening for ``goodness" i.e. the availability of SWEAP observations and the availability of FIELDS observations at a constant cadence within the interval.
\item \textbf{Subsection \ref{app_noise_removal}:}  $\tilde{B}(f)$, the DFT of components of magnetic field time series, \textbf{B}(t) is performed.  Frequencies at which the coherent sharp-peaked noise due to reaction wheels is present are identified, the noise is removed, and the reaction wheel noise-removed \textbf{B}(t) is obtained.

\item \textbf{Subsection \ref{app_downsampling}:}  An anti-aliasing filter is applied and the reaction wheel noise-removed \textbf{B}(t) is downsampled.

\item \textbf{Subsection \ref{app_wavelet_psp}:}  The wavelet energy spectrum of the reaction wheel noise-removed, anti-aliased \textbf{B}(t), $|\tilde{B}(f)|_{psp}^2$ is evaluated. Edge effects in the spectrum due to finite length of time series are reduced.

\item \textbf{Subsection \ref{app_high_f_wave_removal}:}  Energy spectrum of ion-cyclotron frequency coherent waves, $|\tilde{B}(f)|_{wave\_psp}^2$ is identified and removed from $|\tilde{B}(f)|_{psp}^2$, obtaining the energy spectrum of low-frequency turbulent fluctuations, $|\tilde{B}(f)|_{turb\_psp}^2$.
\end{enumerate}

\subsection{PSP/FIELDS}

 The PSP/FIELDS instrument suite (doi: \url{https://hpde.io/NASA/NumericalData/ParkerSolarProbe/FIELDS/index.html}) makes in situ observations of electromagnetic fields (\cite{Bale2016}). Level 2 magnetic field data (in RTN coordinates) from PSP Encounters 1 and 2 observed by the flux gate magnetometer (MAG) are obtained.  The time series data from each encounter is split into continuous non-overlapping intervals of length $\sim$ 15 minutes. During Encounter 1, the sampling rate of MAG varies between  $\sim$ 73.24 Sa/s, 146.48 Sa/s, and 292.96 Sa/s. Intervals are discarded if there is a change in sampling rate within the interval. This removes $\sim$ 26 intervals, representing 2.3$\%$ of initial measurements. During Encounter 2, the magnetic field is sampled at a constant rate of 146.48 Sa/s and no such selection is necessary. The number of data points in each interval varies between $2^{16}$, $2^{17}$, and $2^{18}$ depending on the sampling rate in Encounter 1 and is a constant $2^{17}$ in Encounter 2. 

 Each interval is extended by $\sim$ 7.5 minutes on either side into adjacent intervals resulting in a larger interval of length $\sim$ 30 minutes. The wavelet energy spectrum of the extended interval is evaluated. However, the spectrum corresponding to the extended 7.5 minutes on either side is discarded to remove the cone of influence (COI) effects, making the analyzed intervals effectively non-overlapping.

\subsection{PSP/SWEAP}

 The PSP/SWEAP instrument suite (doi: \url{https://hpde.io/NASA/NumericalData/ParkerSolarProbe/SWEAP/index.html} ) makes in situ observations of solar wind thermal plasma that are processed to evaluate the velocity distribution functions (VDFs) of ions and electrons and associated moments (\cite{Kasper2016}). We use data that are obtained directly or derived from SWEAP observations. Using $w_p$ ($w_p = \sqrt{2k_BT_p/m_p}$), the moment of proton VDF observed by Solar Probe Cup (SPC) (\cite{Case_2020}), the isotropic proton temperature, $T_p$, is evaluated by,
$$T_{p}[eV] = \dfrac{m_p (w_p[m/s])^2}{2e}.$$
Here, $m_p$ = $1.67 \times 10^{-27}$ Kg, $k_B$ = $1.38 \times 10^{-23}$ J/K and e = $1.602 \times 10^{-19}$ J/eV. 
We use parallel  and perpendicular electron temperatures (with respect to the direction of magnetic field), $T_{\parallel_e}$ and $T_{\perp_e}$, measured by \cite{Halekas2020} using SWEAP/SPANe observations (\cite{Whittlesey_2020}) and estimate isotropic electron temperature, $T_e$,  $$T_e [eV] = \dfrac{T_{\parallel_e} [eV] + 2T_{\perp_e} [eV]}{3}.$$
 We also use the number density of protons, $n_p [cm^{-3}]$, components of the solar wind velocity in RTN coordinates, $V_{sw_{R,T,N}} [m/s]$ (\cite{Case_2020}) and its magnitude, $V_{sw} [m/s]$.
 The observations of the components of solar wind velocity and magnetic field are downsampled to a common cadence (1 minute) and the angle between them, $\theta_{BV_{sw}}$ is calculated by,
$$\theta_{BV_{sw}} = \cos^{-1} \bigg(\dfrac{\textbf{B} \cdot \textbf{V}_{sw}}{|\textbf{B}||\textbf{V}_{sw}| }\bigg).$$
The average values of $w_p[m/s]$, $T_p [eV]$, $T_e [eV]$, $n_p [cm^{-3}]$, $V_{sw} [m/s]$, $B [nT]$, and $\theta_{BV_{sw}}$ are obtained for each interval. Intervals where the SWEAP observations are not available are excluded. This excludes 17 intervals in Encounter 1 and 62 intervals in Encounter 2. Based on the availability of FIELDS and SWEAP data, the number of good intervals obtained in Encounter 1 is 1072, and Encounter 2 is 1046.

\section{Evaluation of turbulent energy spectrum from PSP observations}
\label{app_sec_psp_spectrum}

\subsection{Identification and removal of coherent noise due to reaction wheels} \label{app_noise_removal}

\begin{figure}
    \centering
    \plotone{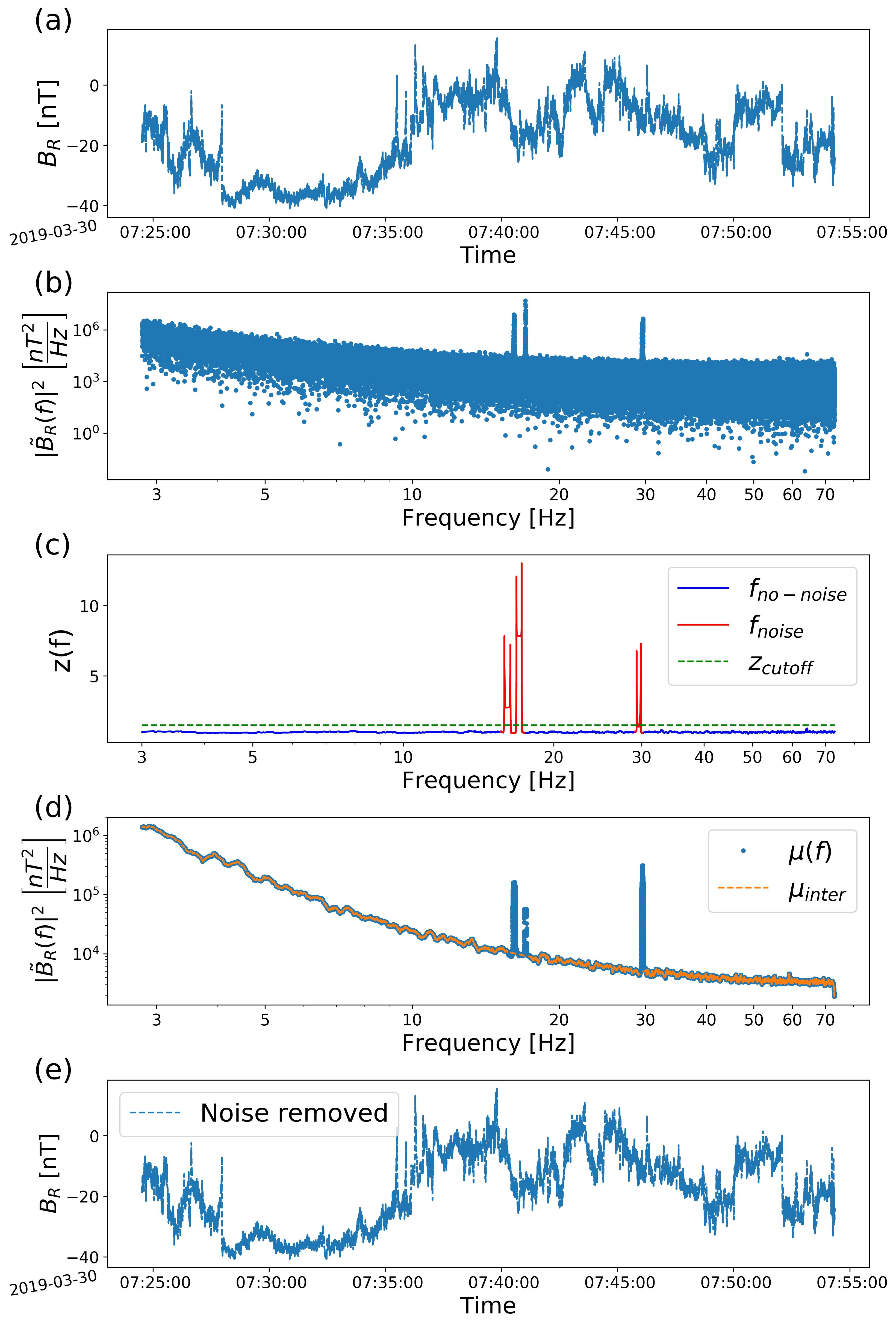} 
    \caption{Reaction wheel noise removal routine for a 30 minute long PSP interval: (a) Time series of R component of magnetic field, $B_R$ and its (b) energy spectrum, $|\tilde{B}_R(f)|^2$  before noise removal. Three sharp noise peaks are evident at 16 Hz, 17 Hz, and 29.6 Hz. (c)The function, z(f) (solid line), and the threshold, $z_{cutoff}$ (green dashed) are plotted against frequency. The frequency region where the noise occurs, $f_{noise}$ (red) is efficiently identified from the frequency region where there is no noise, $f_{no-noise}$ (blue). (d) The evaluated moving window mean of $|\tilde{B}_R(f)|^2$, $\mu (f)$ (blue) and $\mu_{inter} (f)$ (orange)  obtained by considering $\mu (f)$ over $f_{no-noise}$ and interpolating it into $f_{noise}$. (e) The noise removed time series of $B_R$.  }
    \label{fig_noise_removal_routine}
\end{figure}

PSP has reaction wheels to maintain the Sun-facing orientation of the spacecraft during encounters. These wheels generate coherent, sharp-peaked noise at the rotation frequencies of the wheels, as well as harmonic and beat frequencies. The frequency ranges of the reaction wheel noise lie in the dissipation range of the turbulent energy spectrum. The wavelet transform has a broad frequency response, which leads to the leaking of the noise energy to adjacent frequencies, considerably affecting the slope of the energy spectrum in the dissipation range. Hence, in each interval, the reaction wheel noise is identified and removed using a moving-window standard deviation method via the following routine (illustrated in Figure \ref{fig_noise_removal_routine}). \revise{This method is similar to the one used in \cite{Bandyopadhyay_2018}, where the authors employed a Hampel filter to remove the reaction wheel noise.}
 
\subsubsection{Identification of reaction wheel noise}

 We evaluate $\tilde{B}_{i}(f)$, the Fourier transforms of the components of magnetic field time series, $B_i(t)$, and the corresponding energy spectra, $|\tilde{B}_{i}(f)|^2 $, where, \textit{i} represents the components in RTN co\"ordinates. The reaction wheel noise manifests in the energy spectrum at frequencies higher than 2.8 Hz, and thus the energy spectra at these frequencies are processed using the following routine. For energy spectrum of each of the RTN components of magnetic field time series , $|\tilde{B}_{i}(f)|^2 _{f>2.8 Hz} $:

\begin{itemize}

\item The moving window mean, $\mu (f)$ and standard deviation, $\sigma(f)$ is evaluated, with a window of a constant width of 0.4 Hz. The window width is chosen so that it is similar to the width of the noise peaks, so as to efficiently identify the latter.

\item A quantity, $z(f) = \dfrac{\sigma(f)}{\mu(f)}$ is defined whose value spikes at frequencies where the noise peak occurs . This behavior allows $z(f)$ to be used as a criterion to identify the reaction wheel noise.

\item An empirical threshold, $z_{cutoff} = 1.4 \times \left< z(f) \right>$, is defined based on the mean value of $z(f)$ averaged over all frequencies $>$ 2.8 Hz. Minor variations in the cutoff value do not qualitatively affect the results. Frequencies where $z(f)>z_{cutoff}$ are identified as the frequencies where reaction wheel noise is present, $f_{noise_{i}}$.
\end{itemize}
 The noise frequency region encompassing frequencies where noise occurs in the energy spectra of all magnetic field components is defined as  $f_{noise} = f_{noise_R} \cup f_{noise_N} \cup f_{noise_T}$.

\subsubsection{Removal of reaction wheel noise}

  Considering the DFT, $\tilde{B}_{i}(f)$, and the energy spectrum, $|\tilde{B}_{i}(f)|^2 $, of each of the RTN components of the magnetic field time series over the entire range of frequencies in the Fourier frequency domain:
\begin{itemize}
\item The moving window mean of $|\tilde{B}_{i}(f)|^2 $ is evaluated with a window of a constant width of 0.4 Hz,  at frequencies without reaction wheel noise, $f_{no-noise} $ and is interpolated into the frequencies where noise occurs, $ f_{noise} $. Let this interpolated moving window mean be $\mu_{inter}(f)$.

\item The noise-removed energy spectrum, $|\tilde{B}_{i}(f)|^2_{\textit{noise removed}}$ is obtained by retaining the values of $|\tilde{B}_{i}(f)|^2 $ at $ f_{no-noise} $ and replacing with the corresponding values of $\mu_{inter}(f)$ at $f_{noise}$.

\item Magnitude of the noise-removed Fourier transform, $\tilde{B}_{i}(f)_{\textit{noise removed}}$ is evaluated by calculating the square root of $|\tilde{B}_{i}(f)|^2_{\textit{noise removed}}$.

\item Phases of $ \tilde{B}_{i}(f)_{\textit{noise removed}}$ are obtained by retaining the phases of $\tilde{B}_{i}(f)$ at $ f_{no-noise} $ and adding randomized phases at $ f_{noise}$, where there is interpolation.

\item The inverse Fourier transform of $ \tilde{B}_{i}(f)_{\textit{noise removed}}$ is evaluated to obtain the noise-removed magnetic field time series, $B_i(t)_{\textit{noise removed}}$.
\end{itemize}

\subsection{Downsampling magnetic field time series and anti-aliasing} \label{app_downsampling}

    In this work, we evaluate the energy spectrum using FIELDS/MAG observations. We consider the energy spectrum up to 10 Hz as the energy spectrum hits the noise floor at $\sim$ 10 Hz for Encounters 1 and 2 (\cite{Bowen_noise}). The magnetic field time series data is oversampled for our purpose. Therefore, we downsample it to a sampling rate of $\sim$ 36.62 Sa/s (Nyquist frequency of 18.31 Hz) in each interval.
    
    On downsampling, the energy at frequencies (previously present) that are greater than the new Nyquist frequency is folded into the lower frequencies, leading to an erroneous enhancement of energy at the latter. This aliasing causes an unphysical increase in slope towards the very end of the energy spectrum in the dissipation range. Hence, a Butterworth low-pass filter of order 10 and a cut-off frequency of 18.31 Hz is applied before downsampling to avoid aliasing. 

 \subsection{Evaluation of wavelet energy spectrum} \label{app_wavelet_psp}

\subsubsection{Wavelet transform}

 A wavelet transform is employed to evaluate the energy spectral density of magnetic field fluctuations. Wavelet transforms resolve the energy of a signal in both time and frequency (\cite{Terrence_compo}, \cite{Podesta_2009}). The wavelet transform for a time series of a magnetic field component, $B_i(t)$ of length N is written as
\begin{equation}
W_i(s,t) = \sum_{j=0}^{N-1} B_i(t_j) \psi \bigg(\dfrac{t_j - t}{s}\bigg), \hspace{2cm} i=1,2,3.
\end{equation}

Here \textit{s} is the scale (1/frequency) at which the transform is evaluated and $\psi(t/s)$ is the wavelet function, which acts as a window of width \textit{s}. In this work, we use a Morlet wavelet,
\begin{equation}
\psi(\eta) = \pi^{-1/4} e^{i \omega_0 \eta} e^{-\eta^2/2} \hspace{0.75cm} \text{with} \hspace{0.75cm} \omega_0 = 6.
\end{equation}

The wavelet energy spectrum of $B_i(t)$ is evaluated by calculating $|W_i(s,t)|^2$.

Errors occur at the beginning and end of a wavelet transform due to the finite length of time series and the regions where these errors arise are known as the cone of influence (COI). However, these edge effects diminish by a factor of $e^2$ over a decorrelation timescale of $s\sqrt{2}$. The edge effects can be minimized by discarding the data corresponding to the decorrelation length at each frequency at both ends of the wavelet transform, as described in the following section.

\subsubsection{Reducing finite-length errors}

The wavelet transform can be evaluated at specific frequencies of interest that lie in the frequency domain of the Fourier transform. We choose the frequencies at which the wavelet energy spectrum is evaluated such that there exists a sufficient resolution for a good comparison with the cascade model energy spectrum. The frequencies in the DFT frequency domain of the central time series of length 15 minutes are split among 400 logarithmic spaced bins and the median frequencies of the bins are evaluated. However, the bin sizes towards the low frequency end are smaller than the resolution of the DFT frequency domain, resulting in many of these bins being empty. Let $f_{\textit{log\_bin}}$ be the set of median frequencies of the non-empty bins. The length of $f_{\textit{log\_bin}}$ varies between 219, 230, and 244 based on the sampling rate in Encounter 1, and is a constant 230 in Encounter 2.

For each interval, the wavelet transforms of the RTN components of the reaction wheel noise-removed, anti-aliased magnetic field time series are evaluated for extended intervals of length 30 minutes. These transforms are 2-D arrays in time-frequency space evaluated at scales, $s = 1/f_{\textit{log\_bin}}$, and time points where the magnetic field is sampled. The transforms corresponding to the extended 7.5 minutes on either side of the central interval are discarded for all frequencies. This process effectively removes the edge effects at frequencies where the decorrelation time scale is less than or equal to 7.5 minutes i.e. 
\begin{equation}
    s\sqrt{2} = \dfrac{\sqrt{2}}{\text{frequency}} <=  7.5 \times 60 \implies \text{frequency} >=  3.14 \times 10^{-3} Hz.
\end{equation}
  Hence, edge effects  are removed for frequencies greater than $3.14 \times 10^{-3}$ Hz i.e well into the inertial region of the energy spectrum, and is sufficient for our analysis.  The transforms corresponding to frequencies less than  $3.14 \times 10^{-3}$ Hz in $f_{\textit{log\_bin}}$ are discarded to obtain 2-D wavelet transform arrays, $W_{R,T,N}(s,t)$ of time length 15 minutes. The energy spectral density of observed magnetic fluctuations, $|\tilde{B} (f)|^2_{psp}$, is evaluated as
  \begin{equation}
      |\tilde{B} (f)|^2_{psp} = \sum_{t} \bigg( |W_R(s,t)|^2 + |W_T(s,t)|^2 +|W_N(s,t)|^2 \bigg).
  \end{equation}

\subsection{Removal of ion-cyclotron frequency coherent wave energy}
\label{app_high_f_wave_removal}

\begin{figure*}[h]
         \centering
         \plottwo{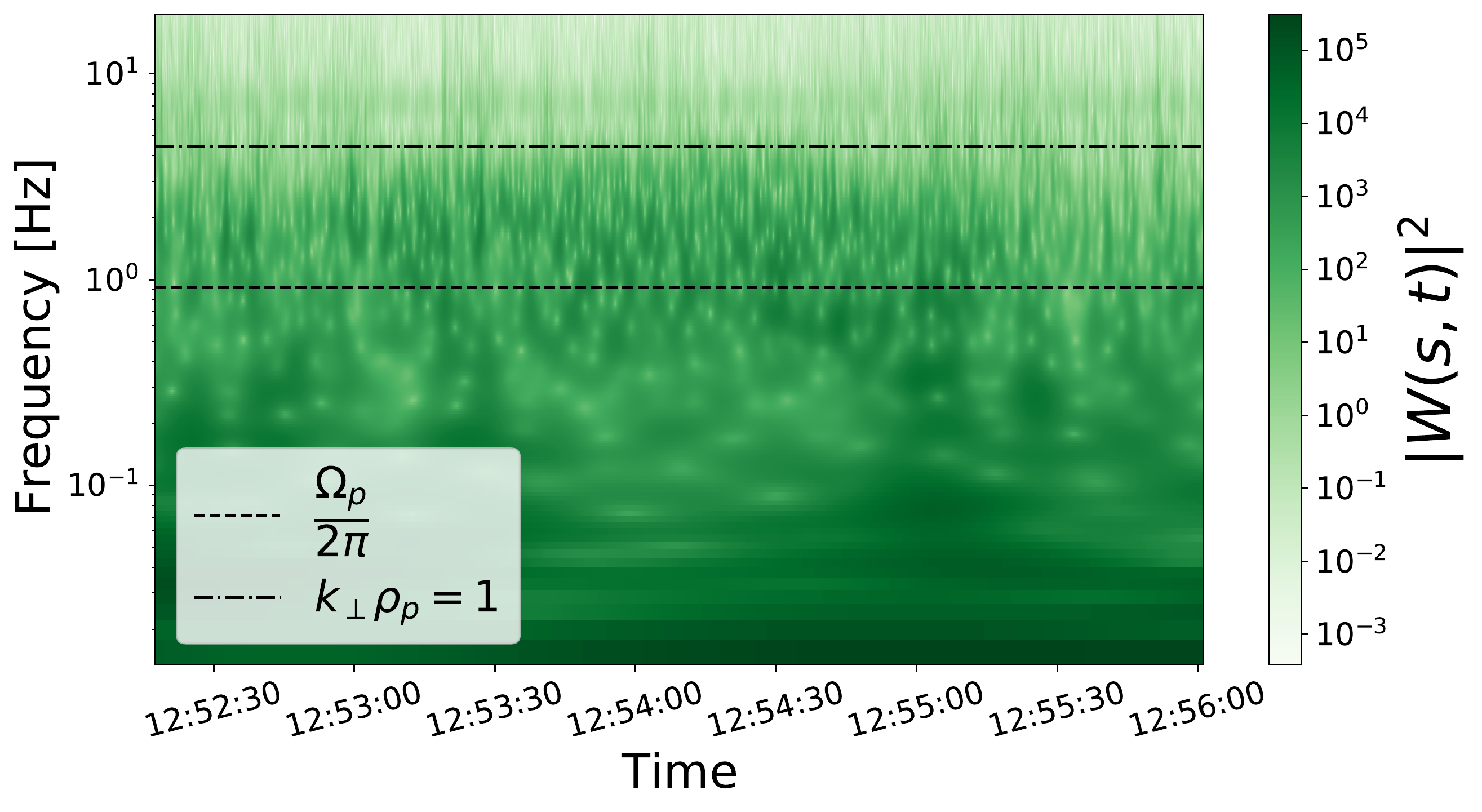}{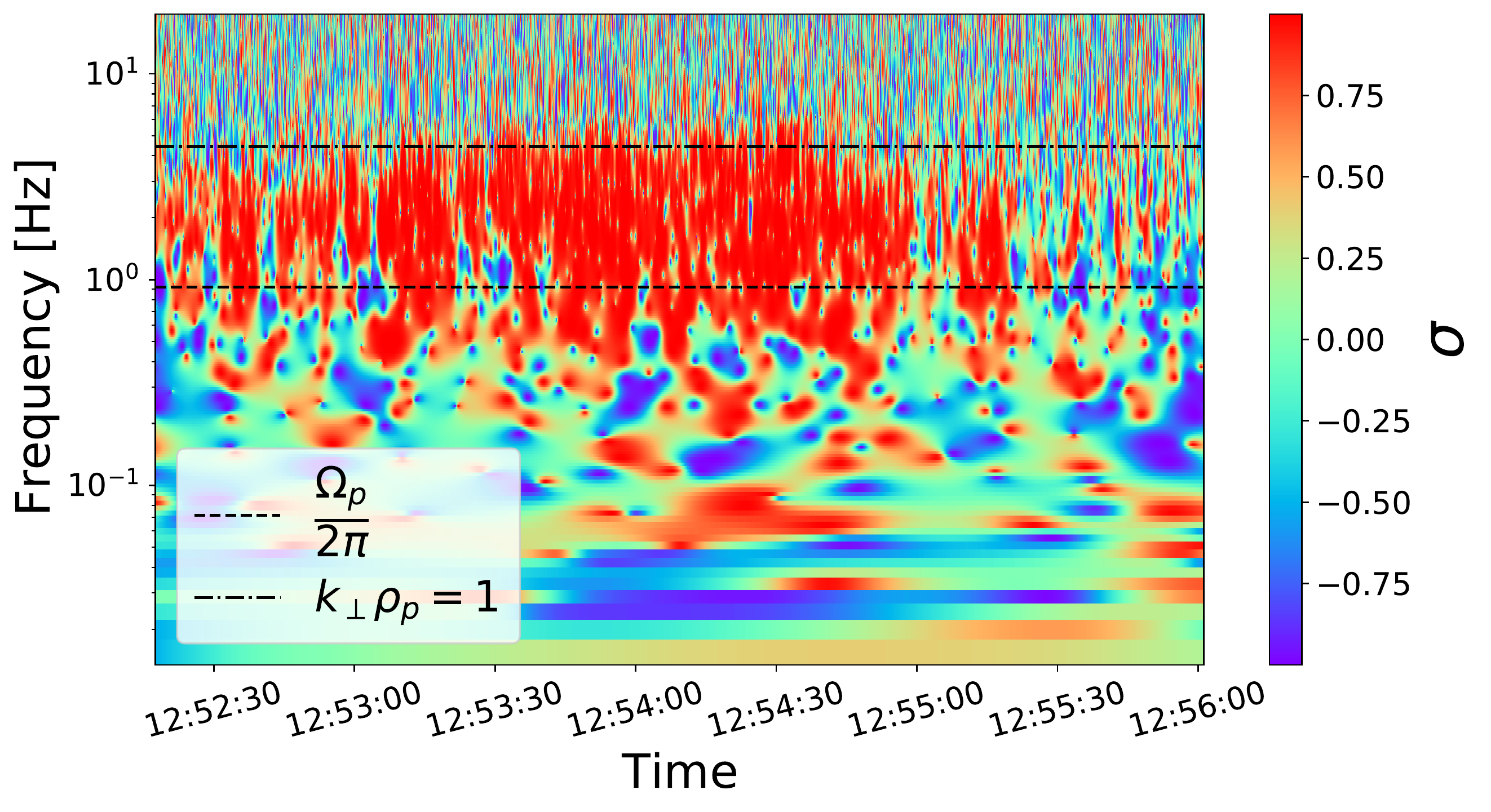}
    \caption{ The wavelet energy spectrum (left panel) and the cross-coherence, $\sigma$ (right panel)  of an interval at 0.21 AU where an ion-cyclotron frequency wave is identified are plotted as functions of time and frequency. We observe an increased energy and a corresponding enhancement in $\sigma$ at frequencies ($\sim$ 2 Hz) near the proton gyrofrequency where the wave occurs.}
    \label{fig_wavelet_energy}
\end{figure*}

Parallel propagating Ion Cyclotron Waves (ICWs) and Fast Magnetosonic/Whistler waves (FM) are often identified near kinetic scales in Encounters 1 and 2 observations (\cite{Bowen_ICW}, \cite{Verniero_2020}). These waves occur most often at frequencies greater than 0.2 Hz. Increased energy is observed at the wave frequencies in the wavelet energy spectrum of intervals where the waves occur (left panel of Figure \ref{fig_wavelet_energy}). The \cite{Howes2008} model assumes a cascade of low-frequency Alfv\'enic turbulent fluctuations and doesn't account for the presence of these ion-cyclotron frequency waves. Hence, it is essential to identify and remove the energy due to these waves from the observed spectra before comparison with the model-estimated spectra.

Both the ICW and FM modes have circular polarization in the plane perpendicular to the local magnetic field direction. There is a high phase coherence between the magnetic field fluctuations along the two perpendicular directions in this plane. This phase coherence is reflected in the corresponding wavelet transforms as well. We can use phase coherence as a criterion to identify these waves. A co\"ordinate system where one of the axes is parallel to the local magnetic field and the other two axes are perpendicular to it is best suited for quantifying this coherence. The local magnetic field direction depends on the scale under consideration, \textit{s}, and so does the co\"ordinate system defined to measure coherence.

\subsubsection{Phase coherence, $\sigma$: criteria to identify wave energy} \label{subsec_cross_coherence}
 
 At each frequency (1/\textit{s}) greater than 0.2 Hz at which wavelet transforms are evaluated, the energy due to ion-cyclotron frequency phase coherent waves is identified via the routine described below, following \cite{Bowen_ICW} and \cite{Lion_2016}:
\begin{itemize}
\item A scale-sensitive co\"ordinate system, XYZ, is evaluated, where $\hat{\textbf{Z}}$ is parallel to the local magnetic field. $\hat{\textbf{X}}$ and $\hat{\textbf{Y}}$ are in the plane perpendicular to $\hat{\textbf{Z}}$. Note that the XYZ co\"ordinate system is time-dependent.
\item The local magnetic field direction, $\hat{\textbf{B}}^s_{local}$, is evaluated in RTN coordinates. 
$$\hat{\textbf{Z}} = \hat{\textbf{B}}^s_{local} = \bigg[ \dfrac{<B_R>_s}{|\mathbf{B}_s|}, \dfrac{<B_T>_s}{|\mathbf{B}_s|}, \dfrac{<B_N>_s}{|\mathbf{B}_s|} \bigg], $$ $$ |\mathbf{B}_s| = \sqrt{<B_R>^2_s + <B_T>^2_s + <B_N>^2_s},$$
where $<>_s$ is the moving window average with a window whose width is equal to the scale \textit{s}.
\item The first axis perpendicular to $\hat{\textbf{Z}}$,  $\hat{\textbf{X}}$ is evaluated by taking the cross product of $\hat{\textbf{Z}}$ and minimum variance direction, $\hat{\textbf{B}}_{MVA}$. The second axis perpendicular to $\hat{\textbf{Z}}$, $\hat{\textbf{Y}}$ is evaluated by taking the cross product of $\hat{\textbf{Z}}$ and $\hat{\textbf{X}}$, forming a right-handed coordinate system, XYZ. 
$$\hat{\textbf{X}} = \hat{\textbf{Z}} \times \hat{\textbf{B}}_{MVA},$$
$$\hat{\textbf{Y}} = \hat{\textbf{Z}} \times \hat{\textbf{X}}.$$

The first of the above cross products of $\hat{\textbf{Z}}$ can be evaluated with an arbitrary vector not parallel to it. Using a different vector in place of $\hat{\textbf{B}}_{MVA}$ would rotate $\hat{\textbf{X}}$ and $\hat{\textbf{Y}}$ by an angle, but the phase coherence is invariant to this rotation.

\item The wavelet spectra in RTN coordinates are transformed to XYZ coordinates.
$$W_{X}(s,t) = \textbf{W}(s,t) \cdot \hat{\textbf{X}}, \hspace{0.75cm} W_{Y}(s,t) = \textbf{W}(s,t) \cdot \hat{\textbf{Y}}, \hspace{0.75cm} W_{Z}(s,t) = \textbf{W}(s,t) \cdot \hat{\textbf{Z}},$$

where $\textbf{W}(s,t) = \big[W_R(s,t), W_N(s,t), W_T(s,t) \big]$ is the wavelet transform vector as a function of time and frequency.

\item In the case of circular polarization, the magnetic field fluctuation components along  $\hat{\textbf{X}}$ and $\hat{\textbf{Y}}$ have a phase lag of $\pi/2$. This lag is reflected in the phases of the corresponding wavelet transforms $W_X$ and $W_Y$  as well. 

\item Cross-coherence, $\sigma$ is defined to quantify a phase coherence in XY plane 
\begin{equation}
 \sigma (s,t) = \dfrac{2 \times \text{Im}\big[W_X(s,t)W_Y^*(s,t)\big]}{|W_X(s,t)|^2 + |W_Y(s,t)|^2}.
\end{equation}

\end{itemize}
Upon repeating the above routine for each frequency (1/\textit{s}) of wavelet transform, a 2-D array, $ \sigma (s,t) $ is obtained. Increased values of $|\sigma|$ are observed at wave frequencies corresponding to the observed increased energy in the wavelet energy spectrum (right panel of Figure \ref{fig_wavelet_energy}), confirming the presence of ion-cyclotron frequency phase coherent waves.

\begin{figure}[h]
    
    \begin{minipage}{0.5\textwidth}
        \centering
         \includegraphics[width=7cm]{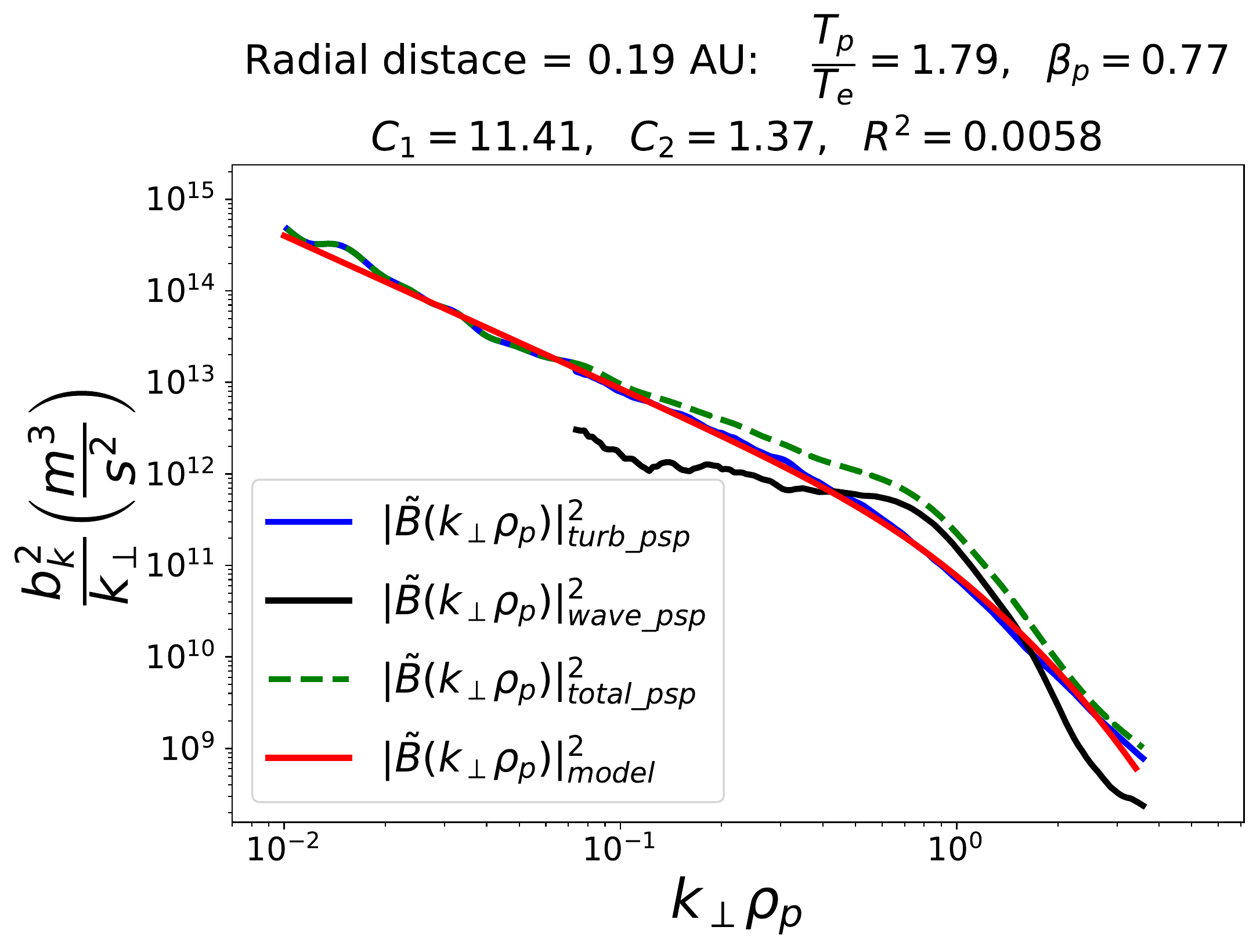} 
    \end{minipage}
           \begin{minipage}{0.5\textwidth}
            \centering
            \includegraphics[width=7cm]{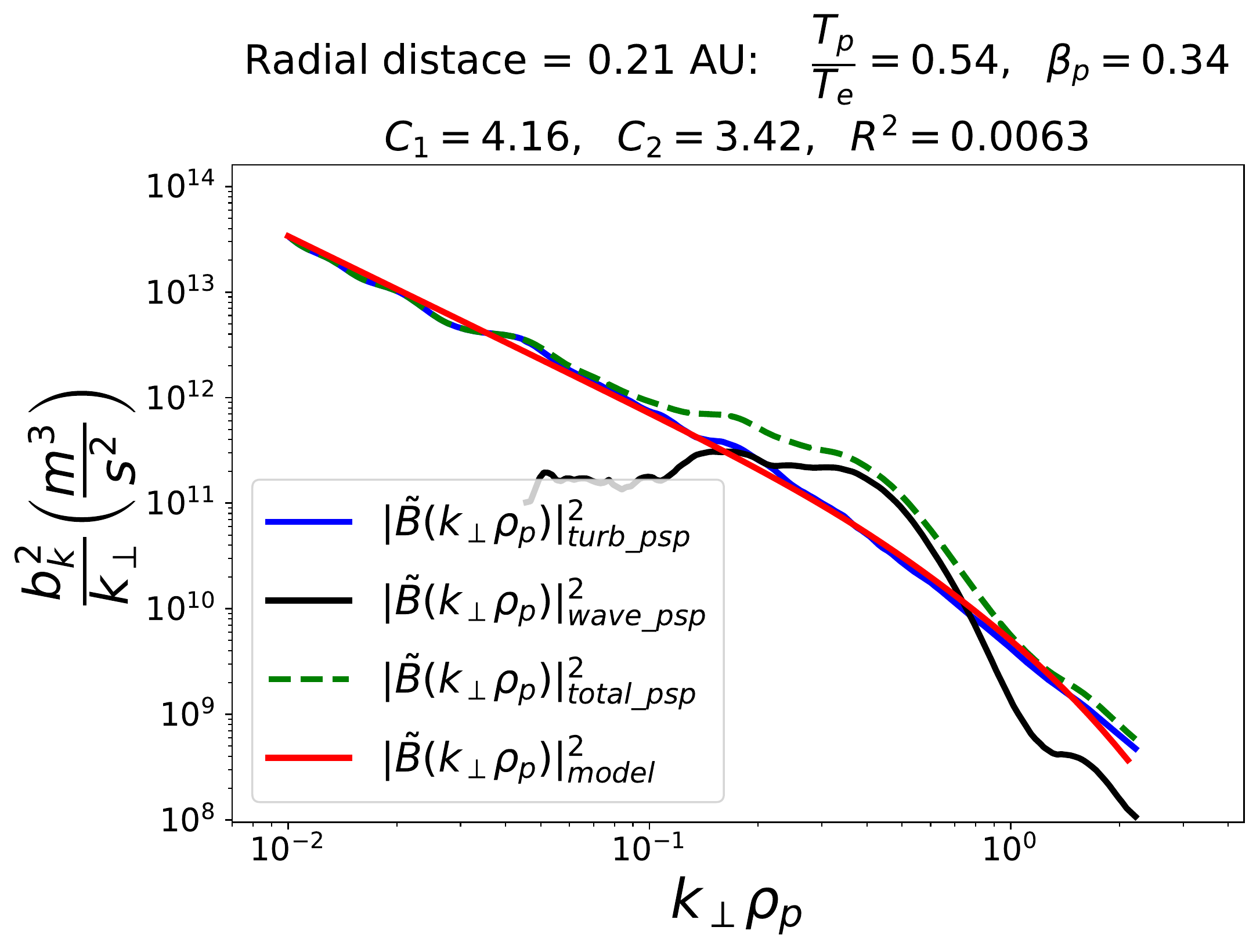}
            \end{minipage}
          
          \caption{Identification and removal of ion-cyclotron frequency wave energy in two intervals where the resulting turbulent energy spectrum is well described by the \cite{Howes2008} model. The interval on the right panel is considered in Figure \ref{fig_wavelet_energy}.}
          \label{fig_wave_energy_removal}
\end{figure}

 Motivated from previous works using 1 AU data (\cite{Lion_2016}) and PSP observations (\cite{Bowen_ICW}), a threshold of 0.7 on the value of $|\sigma|$ is set to identify the presence of ion-cyclotron frequency coherent waves. For a given interval, the energy at the frequency-and-time points where $|\sigma(s,t)|> 0.7$ is designated as coherent, while the energy at the frequency-and-time points where $|\sigma(s,t)|\leq 0.7$ are designated incoherent. We sum over all time values in the interval, resulting in a frequency spectrum of the coherent $|\tilde{B}(f)|^2_{wave\_psp}$ and incoherent $|\tilde{B}(f)|^2_{turb\_psp}$ energy (Figure \ref{fig_wave_energy_removal}).

\begin{align} \label{psp_psd_2}
& |\tilde{B} (f)|^2_{wave\_psp} = \sum_{t_{coherent}} \left( |W_X(s,t)|^2 + |W_Y(s,t)|^2 +|W_Z(s,t)|^2 \right) \nonumber\\
& |\tilde{B} (f)|^2_{turb\_psp}  = \sum_{t_{non-coherent}} \left( |W_X(s,t)|^2 + |W_Y(s,t)|^2 +|W_Z(s,t)|^2 \right) = |\tilde{B} (f)|^2_{psp} - |\tilde{B} (f)|^2_{wave\_psp}
\end{align}

 In the plasma rest frame, a $\sigma$ value of 1 would imply electron-resonant or right-handed waves (Whistlers) and -1 would imply ion-resonant or left-handed waves (ICWs). The magnetic field and wavelet transform are measured in the spacecraft frame. The following equation dictates the transformation from spacecraft frame to plasma frame,
 \begin{equation} \label{psp_psd_1}
\omega_{spacecraft} = \omega_{plasma} + \textbf{k} \cdot \mathbf{V_{sw}}.  
 \end{equation}

As this work focuses on modeling the low-frequency fluctuations, we leave a detailed exploration of the ion-cyclotron-frequency waves to future studies.

\subsubsection{Verification of accuracy of the $\sigma$ criteria} \label{sect_polarization}

\begin{figure}[h]
 \begin{minipage}{0.5 \linewidth}
     \centering
     \includegraphics[width=7cm, height=2.5cm]{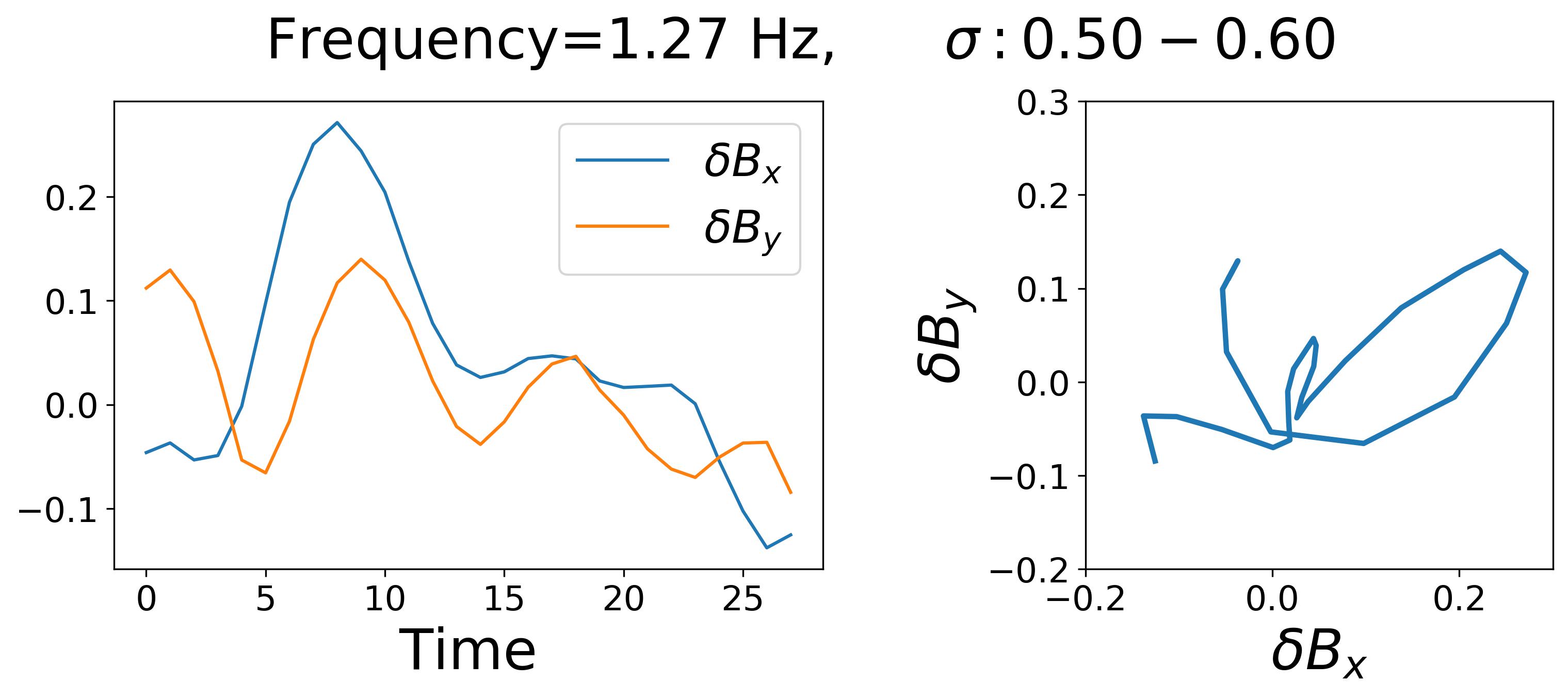}
     
     \includegraphics[width=7cm, height=2.5cm]{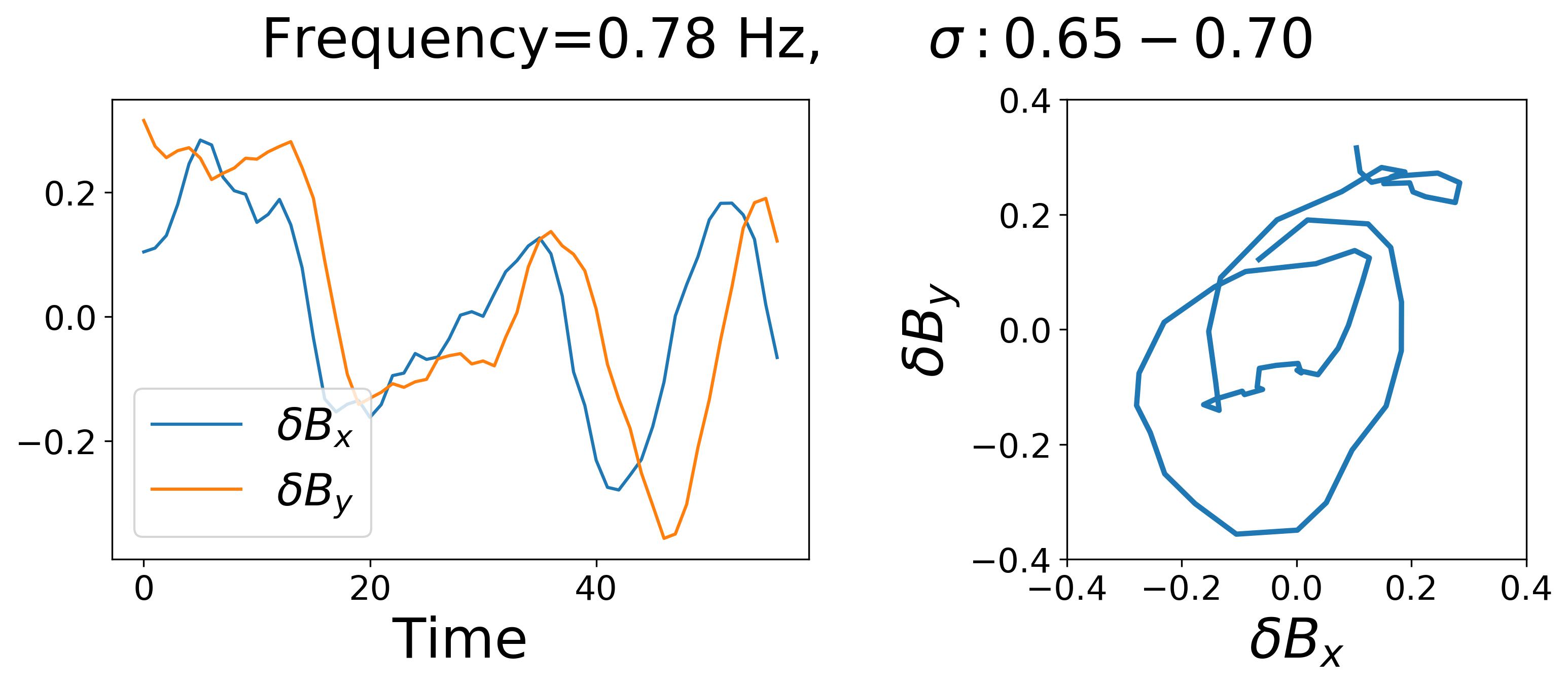}
     
     \includegraphics[width=7cm, height=2.5cm]{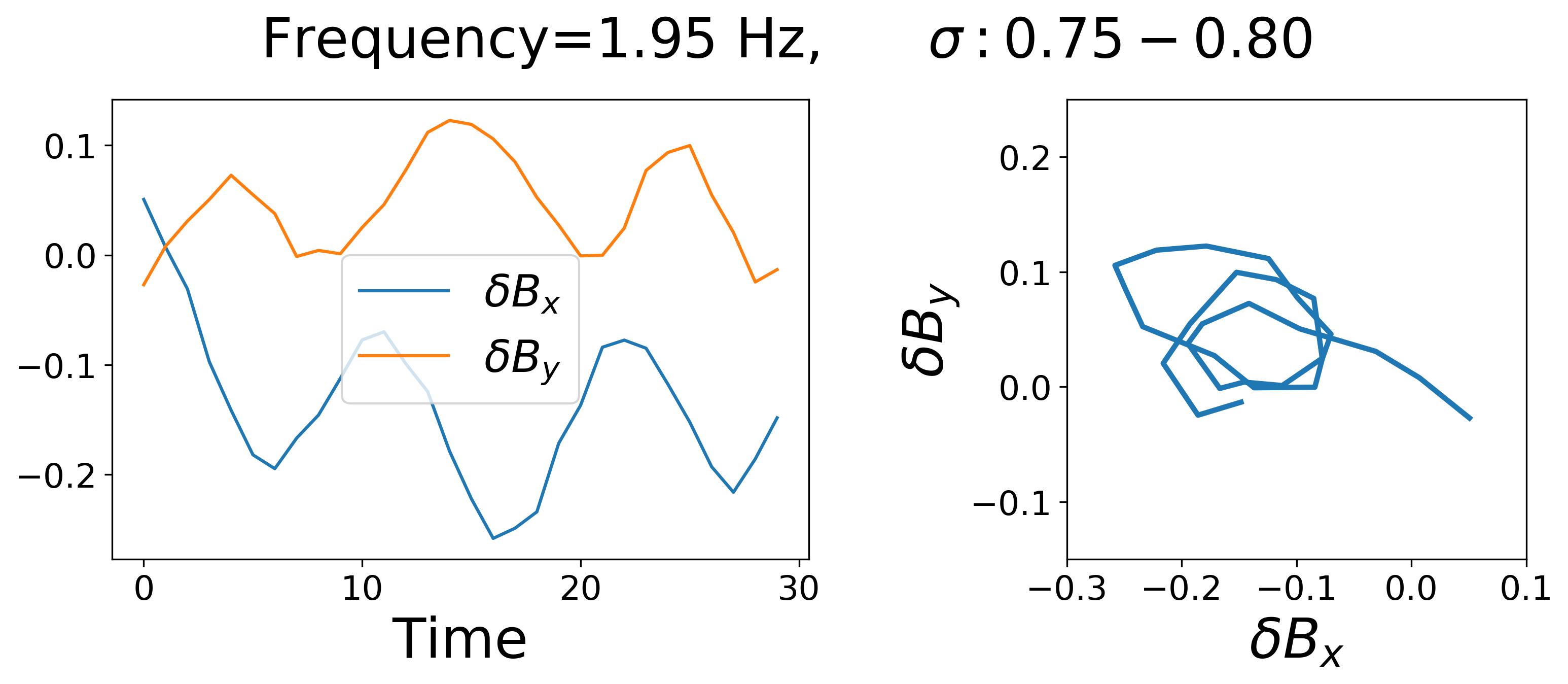}
     
     \includegraphics[width=7cm, height=2.5cm]{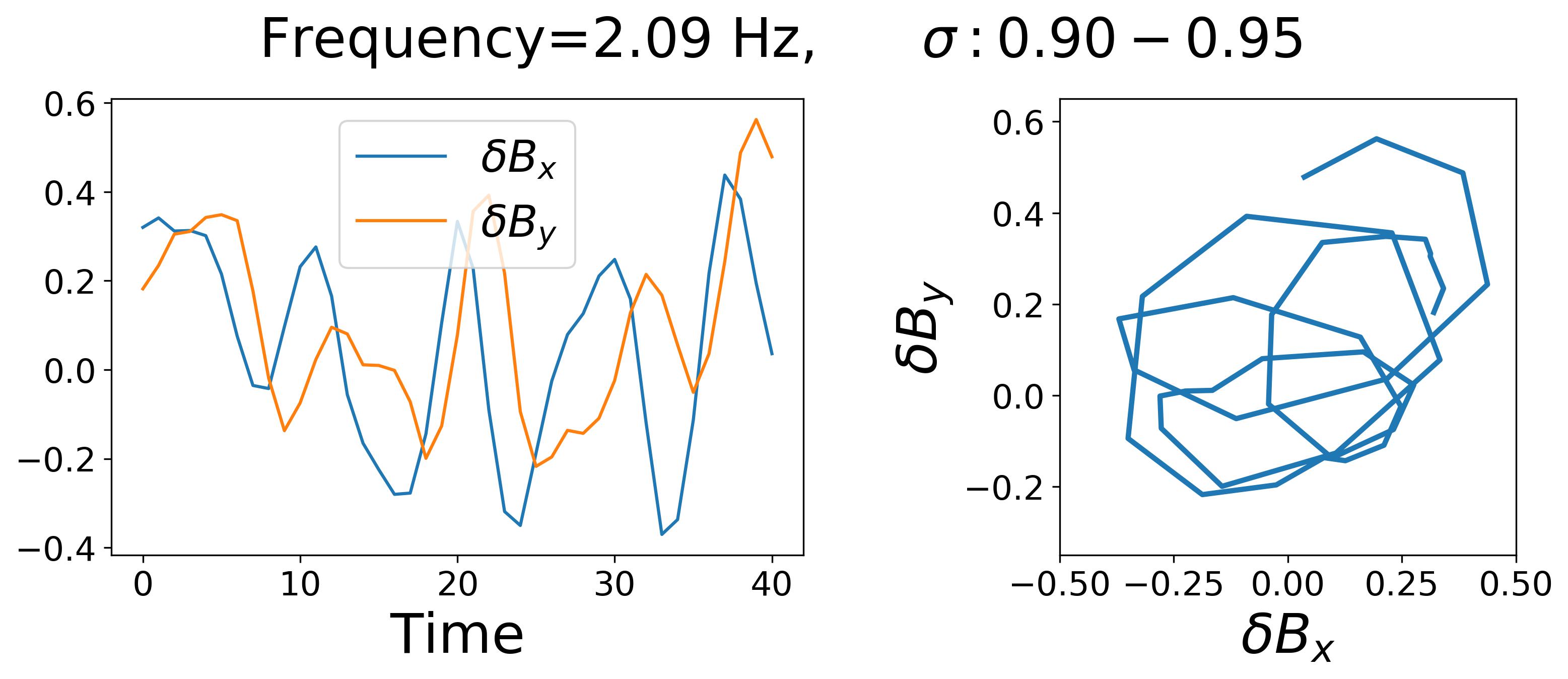}
     \end{minipage}
        \begin{minipage}{0.5 \linewidth}
        \centering
        \includegraphics[width=7cm, height=2.5cm]{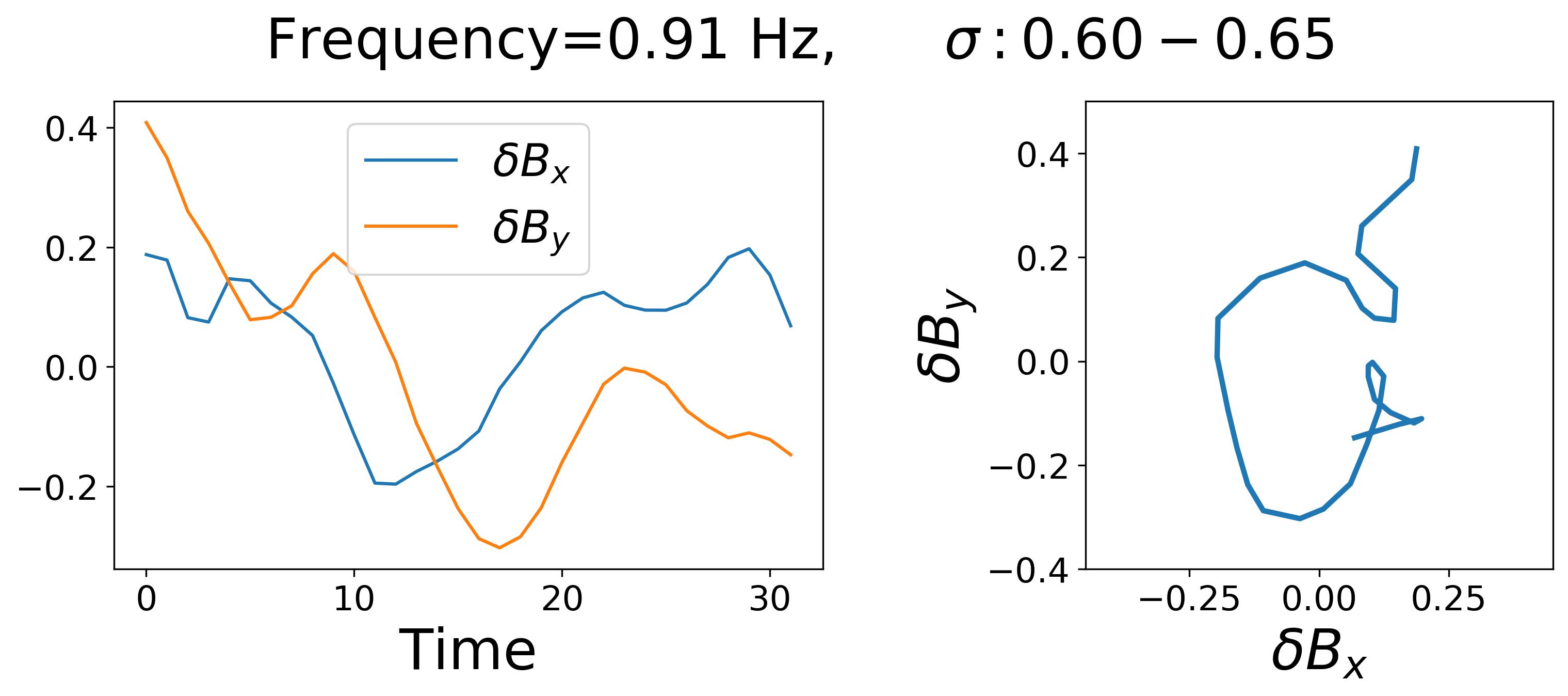}
        
        \includegraphics[width=7cm, height=2.5cm]{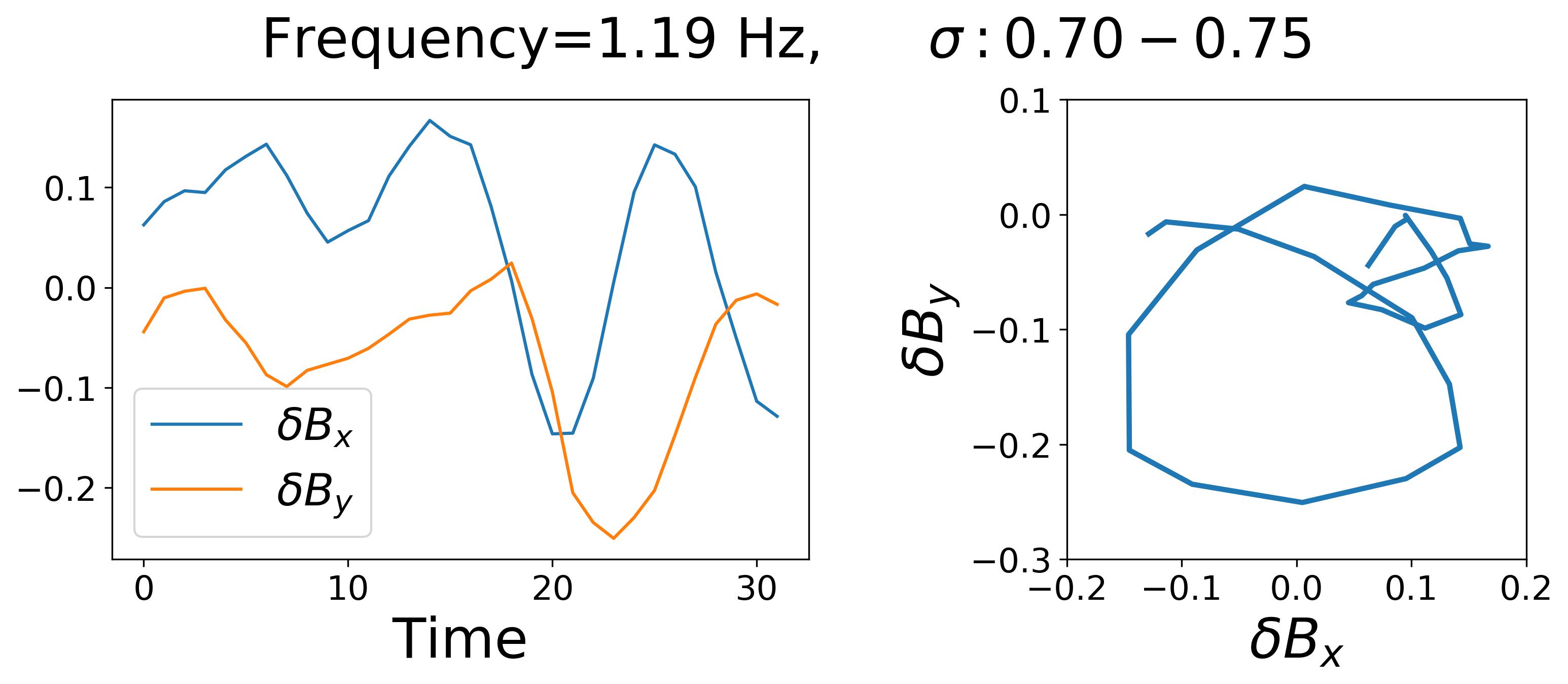}
        
        \includegraphics[width=7cm, height=2.5cm]{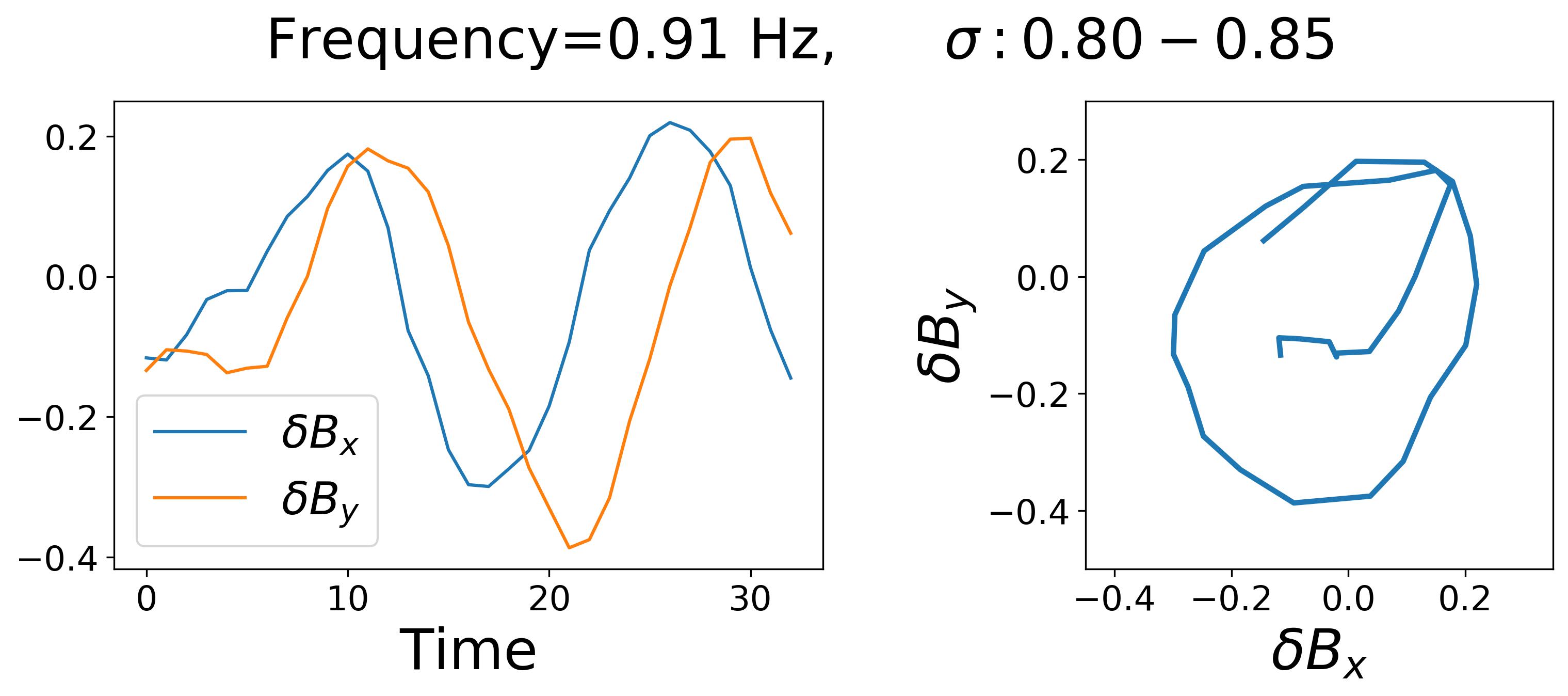}
        
        \includegraphics[width=7cm, height=2.5cm]{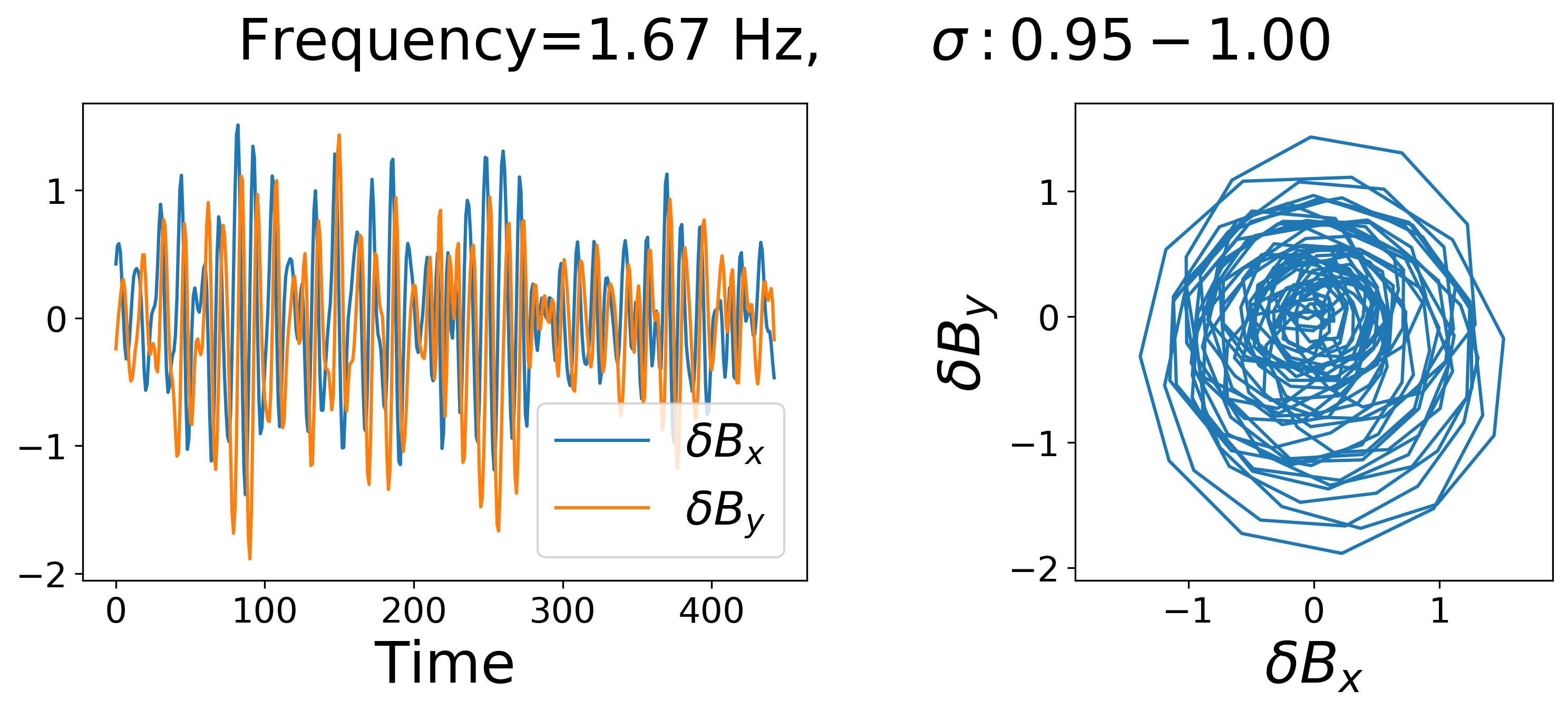}
         \end{minipage}
       
     \caption{Polarization observed in the plane perpendicular to the local magnetic field in selected intervals with various ranges of $\sigma$ values.}   
     \label{fig_coherence}  
 \end{figure}

 Figure \ref{fig_coherence} shows the hodograms of magnetic field fluctuations along $\hat{\textbf{X}}$ and $\hat{\textbf{Y}}$ directions, $\delta B_X (t)$ and $\delta B_Y (t)$, plotted for time intervals with $|\sigma|$ values varying between 0.5 and 1. Circular polarization is evident for $|\sigma| > 0.7$. A threshold of 0.7 is thus inferred to be effective.

 $\delta B_X$ and $\delta B_Y$ are evaluated by a prescription described in \cite{Lion_2016}. The frequencies, $f_{min}$ and $f_{max}$ between which a particular range of coherence occurs in a time series are identified and $$\delta B_{X,Y} = <B_{X,Y}>_{\tau_{max}} - <B_{X,Y}>_{\tau_{min}}.$$  Here,
      $<>_{\tau_{max}}$ is the moving window average with a window of width $1/f_{max}$ which averages out variations with frequencies greater than $f_{max}$. 
      Similarly, $<>_{\tau_{min}}$ averages out variations with frequencies greater than $f_{min}$. The difference $<>_{\tau_{max}} - <>_{\tau_{min}}$ selects the fluctuations with frequencies where the considered coherence is observed.

 \subsection{Normalization of PSP energy spectrum}

   We assume Taylor's hypothesis (\cite{Taylor1938}, \cite{Fredricks_Coriniti}) i.e the time scales of local plasma variations are much longer than the time scale for advection of plasma with respect to the spacecraft and write equation \ref{psp_psd_1} as, $2 \pi f_{spacecraft} =  \textbf{k} \cdot \textbf{V}_{sw}. $
 
  For an anisotropic distribution of energy ($k_{\perp} >> k_{\parallel}$), Taylor's hypothesis is written as $$2 \pi f_{spacecraft} =  \textbf{k}_{\perp} \cdot \textbf{V}_{sw} =  k_{\perp} V_{sw} \cos \theta_{k_{\perp}V_{sw}}.$$
  
  Here, $\textbf{k}_{\perp}$ is perpendicular to $\textbf{B}$ and $\theta_{k_{\perp}V_{sw}}$ = $90^{\circ}$ - $\theta_{BV_{sw}}.$ Hence, Taylor's hypothesis is written as
  $$2 \pi f_{spacecraft} =   k_{\perp} V_{sw} \sin \theta_{B V_{sw}}.$$
 For each interval, the transformation from frequency to perpendicular wavevector space is written in terms of interval-averaged (over 15 minutes) quantities  as
 \begin{equation} \label{psp_psd_3}
 f = \dfrac{k_{\perp} V_{sw} \sin \theta_{BV_{sw}}}{2 \pi}.
 \end{equation}
  The energy spectrum of turbulent fluctuations, $|\tilde{B} (f)|^2_{turb\_psp}$ [$T^2$/Hz]  obtained from equation \ref{psp_psd_2}  is transformed to wavevector space, $|\tilde{B} (k_\perp)|^2_{turb\_psp}$ [$T^2$m]  by equation \ref{psp_psd_3} . Turbulent magnetic field fluctuations are transformed to velocity units by, $\delta B$ [m/s] = $\dfrac{ \delta B [T]}{\sqrt{\mu_0 n_p [m^{-3}] m_p}}$ producing the normalized energy spectrum from PSP observations, $|\tilde{B} (k_\perp)|^2_{turb\_psp}$ [$m^3/s^2$]. Here $\mu_0 = 4 \pi \times 10^{-7} $m kg $s^{-2} A^{-2}$ is the permeability of free space.

\section{Evaluation of energy spectrum from the model } \label{app_model_spectrum}

\subsection{Initial parameters}
 
 For each interval, the \cite{Howes2008} cascade model uses as input, the parameters: $T_{p}/T_{e}$,  $\beta_p$, $w_p/c$ and Kolmogorov constants $C_1$ and $C_2$ and calculates linear gyrokinetic frequencies and damping rates as a function of $k_\perp \rho_p$. Here
 $$\beta_p = \dfrac{n_p[cm^{-3}] \times T_p [eV] \times 1.602 \times 10^{-12}}{(B^2[T^2] \big / 8 \pi) \times 10^8} $$ is the ratio of proton thermal pressure to magnetic pressure. The model then solves the Batchelor's equation (equation \ref{eq_intro_2}, \cite{Batchelor1953}) over a wide range of perpendicular wavenumbers from the lowest, $k_{\perp_i} \rho_p = 10^{-2} $ to the highest, $k_{\perp_f} \rho_e = 1.2 $, following the cascade from the inertial range to electron kinetic scales. Here, $k_0 \rho_p = 10^{-4} $ is the isotropic driving wavenumber.

\subsection{Normalization of model energy spectrum}

The steady-state magnetic field fluctuations as a function of $k_\perp$, $b_k$, are obtained by solving the Batchelor's equation, with the numerical solution evaluating the dimensionless energy spectrum, $\dfrac{1}{k_{\perp} \rho_p}\bigg(\dfrac{b_{k}}{b_{k_i}} \bigg)^2$. Here $b_{k_i}$ is the magnitude of magnetic field fluctuations at the largest scale, $k_{\perp_i}$. The value of $b_{k_i}$ for each interval is constrained using the observed values of plasma parameters as described in Appendix \ref{sect_norm} (equation \ref{norm10}), 
\begin{equation}
 b_{k_i} = C_2^{-1} v_A \bigg(\dfrac{k_0}{k_{\perp i}}\bigg)^{1/3}.
\end{equation}
 The energy spectrum is normalized to physical units [$m^3/s^2$] as
 \begin{equation}
     |\tilde{B}|^2_{model} (k_{\perp},C_1,C_2) [m^3/s^2] = \dfrac{1}{k_{\perp} \rho_p}\bigg(\dfrac{b_{k}}{b_{k_i}} \bigg)^2 \times C_2^{-2}  v_A^2 \bigg(\dfrac{k_0}{k_{\perp_i}}\bigg)^{2/3}  \rho_p ,
 \end{equation}

where, $\dfrac{1}{k_{\perp} \rho_p}\bigg(\dfrac{b_{k}}{b_{k_i}} \bigg)^2$ is evaluated in the code. The mean values of ion gyroradius and Alfv\'en speed  over the interval are evaluated using interval-averaged quantities, $$\rho_p [m] = \sqrt{\dfrac{2 T_p [eV] e}{m_p}} \Biggr / \dfrac{e B [T]}{m_p}, \hspace{1cm} v_A [m/s] = \dfrac{B[T]}{\sqrt{\mu_0 m_p n_p [m^{-3}]}}.$$

\section{Constraining $C_1$ and $C_2$} \label{app_C1_C2_constrain}

For each interval, the values of Kolmogorov constants $C_1$ and $C_2$ are constrained by maximizing the agreement between the energy spectra obtained from PSP observations and the cascade model. The agreement between the spectra is quantified by a function, $R^2$ (Equation \ref{Eq_R2}).
The function $R^2$ is minimized with respect to $C_1$ and $C_2$ to find the best fit of the model energy spectrum to the energy spectrum evaluated from PSP observations. Initially, $R^2$ is evaluated over a dense grid of 391 $(C_1,C_2)$ values where the magnitude of $C_1$ and $C_2$ varies logarithmically from $10^{-1}$ to $10$. A set of local minima of $R^2$ are identified in the $C_1$-$C_2$ parameter plane. All the local minima are then refined until they converge ($R^2_{n+1} - R^2_{n} <= 5 \times 10^{-6}$) using the Levenberg-Marquardt minimization technique. The refined local minimum with the least $R^2$ is the absolute minimum with the corresponding values of $(C_1,C_2)$ being $(C_{1_{best\_fit}},C_{2_{best\_fit}})$. 

The order of magnitude of turbulent energy varies over a wide range from the inertial to dissipation scales. The magnitude of $R^2$ is more sensitive to the inertial scales compared to the dissipation scales. In order to better quantify the difference between $|\tilde{B}|^2_{turb\_psp}$ and  $|\tilde{B}|^2_{model}$ in dissipation scales, we define a second goodness-of-fit metric that focuses specifically on the scales where the proton and electron damping occurs, 

\begin{equation}
\frac{\Delta E}{E}_{diss} = \frac{\sum_{(k_\perp \rho_p)_{diss}} | |\tilde{B}|^2_{turb\_psp} - |\tilde{B}|^2_{model} | \Delta (k_\perp \rho_p) }{\sum_{(k_\perp \rho_p)_{diss}} |\tilde{B}|^2_{turb\_psp} \Delta (k_\perp \rho_p) },
\end{equation}
Here we identify the dissipation region, $(k_\perp \rho_p)_{diss}$ as the region between the scale where the spectral index of $|\tilde{B}|^2_{model}$ hits -2.5 and the scale corresponding to the highest frequency considered (10 Hz).
In intervals where $R^2_{best\_fit} \le 0.03$ and $\frac{\Delta E}{E}_{diss} \le 0.25$, the turbulent cascade is considered to be well described by the \cite{Howes2008} model.

\subsection{Levenberg-Marquardt Algorithm}

The Levenberg-Marquardt algorithm (\cite{Levenberg_1944} and \cite{Marquardt_1963}), which solves the non-linear least squares problem, is employed to minimize $R^2(C_1, C_2)$ once local minima are identified on the initial grid in $(C_1,C_2)$ space. At each step in the minimization routine, it is necessary to evaluate $R^2$ at least at 5 neighboring points around ($C_1,C_2$) to evaluate the first and the second order derivatives of $R^2$ in the $C_1$-$C_2$ plane. Hence, $R^2$ is evaluated at $(C_1+h, C_2)$, $(C_1-h, C_2)$, $(C_1, C_2+h)$, $(C_1, C_2-h)$ and $(C_1 +h, C_2+h)$ at each step. Here h is the step size and its value is chosen to be $5 \times 10^{-4}$.

\section{Estimation of $Q_p$ and $Q_e$} \label{app_normalization}
 
 The proton and electron heating rates are estimated using the linear gyrokinetic damping rates and the magnitudes of steady-state magnetic field fluctuations obtained by solving Batchelor's equation.
  In \cite{Howes2008} model, the energy cascade rate, $\epsilon$ is written as \begin{equation} \label{norm1}
\epsilon(k_\perp)=C_1^{-3/2} k_\perp v_k b_k^2, 
\end{equation} 

where  $b_k = \sqrt{\delta B_{\perp}^2(k_\perp) / 4 \pi n_p m_p}$ and $v_k = v_\perp (k_\perp)$ are the magnitude of magnetic field fluctuations in velocity units and the electron fluid velocity fluctuations respectively and are perpendicular to the mean magnetic field.  Kinetic theory relates velocity and magnetic fluctuations via,
\begin{equation}\label{norm2}
\begin{gathered}
    v_k = \pm \alpha(k_\perp) b_k,\\
    \alpha(k_\perp) = 
     \begin{cases}
      1, & k_\perp \rho_p \ll 1 \\
      k_\perp \rho_p/\sqrt{\beta_p + 2/(1+ T_e/T_p)}, & k_\perp \rho_p \gg 1.
    \end{cases}
\end{gathered}
\end{equation}

For a critically balanced cascade, linear and non-linear frequencies can be equated,
\begin{equation}\label{norm3}
\omega_{nl}(k_\perp)= \omega = C_2 k_{\perp} v_k.
\end{equation}
Normalized gyrokinetic linear frequencies and damping rates are written as 
\begin{subequations}
\begin{equation}\label{norm4.1}
\omega = \pm \overline{\omega}(k_\perp) k_{\parallel} v_A,
\end{equation}

\begin{equation} \label{norm4.2}
\gamma = \overline{\gamma}(k_\perp) k_{\parallel} v_A.
\end{equation}
\end{subequations}
Kinetic theory dictates, $\overline{\omega}(k_\perp) = \alpha(k_\perp)$ at asymptotic fluctuation scale ranges $k_\perp \rho_p \ll 1$ and $k_\perp \rho_p \gg 1$. It's sufficient for us to assume, for all fluctuation scales,
\begin{equation}\label{norm5}
\overline{\omega}(k_\perp) = \alpha(k_\perp).
\end{equation}

\subsection{Expression for heating rate}

Using the steady-state values of $b_k$($k_\perp$) obtained by solving Batchelor's equation, heating rate per mass, Q is written as the integral of heating at all wavenumbers considered in the equation,
\begin{equation} \label{norm6}
Q= \int_{k_{\perp i}}^{k_{\perp f}} d \overline{Q}(k_\perp) = \sum_{k_\perp} \delta \big(\overline{Q}(k_\perp)\big) =   \sum_{k_\perp} \delta(2 \gamma b_k^2 ) .
\end{equation}
Here $\delta \big(\overline{Q}(k_\perp)\big)$ is the heating at each wavenumber. Q has units of $m^2 s^{-3}$ (W/kg in SI units) when $b_k$ has units of m/s. Batchelor's equation is solved in the code with $b_k$ in dimensionless units. Q is normalized to physical units via $b_{ki}$, the value of $b_k$ at  $k_{\perp i}$, the largest turbulent fluctuation scale considered in the model. Using equations \ref{norm2}, \ref{norm3}, \ref{norm4.1}, \ref{norm4.2} and \ref{norm5}, equation
\ref{norm6} is written as
\begin{equation} \label{norm7}
  \delta \big(\overline{Q}(k_\perp)\big) =  \dfrac{2 C_2 \overline{\gamma}(k_\perp) \overline{k}_\perp b_k^3}{ \rho_p} \delta(\text{ln} \overline{k}_\perp),   
\end{equation}
where $\overline{k}_\perp = k_\perp \rho_p$ is a dimensionless parameter.
 Gyrokinetic Landau damping rates, $\overline{\gamma}(k_\perp)$, are decomposed into individual proton and electron damping rates, $\overline{\gamma}_{p,e}(k_\perp)$ (\cite{Howes_2006}). Equation \ref{norm7} is multiplied and divided with $b_{ki}^{3}$ and summed over all wavenumbers,
\begin{equation} \label{norm8}
    Q_{p,e} = \underbrace{\Bigg[ \sum_n 2 C_2 \overline{\gamma}_{n_{p,e}}(k_\perp) \overline{k}_{\perp n}  \bigg(\dfrac{b_{kn}^{3}}{b_{ki}^{3}}\bigg) \delta(\text{ln} \overline{k}_{\perp n}) \Bigg]}_{\text{evaluated from the code}} \dfrac{b_{ki}^{3}}{\rho_p}.
\end{equation}
For each interval, $b_{ki}^3$ is constrained using the interval-averaged observed values of plasma parameters and the best-fit Kolmogorov constants, thus, normalizing $Q_{p,e}$.

\subsection{Normalization of heating rate} \label{sect_norm}

   Using equations \ref{norm1}, \ref{norm2}, \ref{norm3}, \ref{norm4.1} and \ref{norm5}, the energy cascade rate is written as $$\epsilon(k_\perp)=C_1^{-3/2} k_\perp \overline{\omega} \Bigg( \dfrac{ k_{\parallel}^3 v_A^3}{C_2^3 k_{\perp}^3} \Bigg).$$
     At the driving scale, turbulence is isotropic, $k_{\perp}=k_{\parallel}=k_0$ and $\overline{\omega} = 1$.
Hence, the input energy cascade rate at the driving scale, $\epsilon_0$ is written as 
\begin{equation}\label{norm9}
\epsilon_0 = C_1^{-3/2} C_2^{-3} k_0 v_A^3.
\end{equation}
Using equations \ref{norm1} and \ref{norm2}, the magnetic field fluctuations are written as $b_k^3 = \dfrac{\epsilon  C_1^{3/2}}{k_{\perp} \alpha (k_{\perp})} $. A reasonable assumption is made that the dissipation occurred until the cascade reaches $k_{\perp i}$ (largest scale considered for an interval) is negligible, implying  $\epsilon (k_{\perp i}) = \epsilon_0$. Substituting  $\alpha (k_{\perp i}) = 1$,  the value of $b_k^3$ at $k_{\perp i}$ is estimated  as
\begin{equation} 
    b_{ki}^3 = \dfrac{\epsilon_0  C_1^{3/2}}{k_{\perp i}}.
\end{equation}
Using equation \ref{norm9} and substituting for $\epsilon_0$,

\begin{equation} \label{norm10}
b_{k i}^3 = C_2^{-3}  v_A^3 \bigg(\dfrac{k_0}{k_{\perp i}}\bigg).
\end{equation}
 Using equation \ref{norm10} and substituting for $b_{k i}^3$ in equation \ref{norm8}, the normalized heating rates are written as
\begin{equation}\label{norm11}
Q_{p,e} = \Bigg[ \sum_n 2 C_{2_{best\_fit}} \overline{\gamma}_{n_{p,e}}(k_\perp) \overline{k}_{\perp n}  \bigg(\dfrac{b_{kn}^{3}}{b_{ki}^{3}}\bigg) \delta(\text{ln} \overline{k}_{\perp n}) \Bigg] \bigg(\dfrac{k_0}{k_{\perp i}}\bigg) \dfrac{C_{2_{best\_fit}}^{-3}  v_A^3}{\rho_p}.
\end{equation}

\bibliographystyle{apj}
\bibliography{main.bib}
\end{document}